\pdfoutput=1
\documentclass[prd,a4paper,eqsecnum,nofootinbib,preprint,floatfix]{revtex4}
\usepackage{colordvi}
\usepackage{graphicx}
\usepackage{bm}
\usepackage{multirow}
\usepackage{enumerate}
\reversemarginpar

\def\<{\langle}
\def\>{\rangle}

\newcommand{\text}{\rm}
\def\Tr{{\rm Tr}\,}
\def\tr{{\text tr}\,}
\def\Eq#1{Eq.~(\ref{#1})}

\renewcommand\Re{{\text Re}\,}

\def\cO{{\cal O}}

\def\QEK{{\rm QEK}}

\begin{document}

\vspace*{0.5in}

\title{
Breakdown of large-$N$ quenched reduction \\ in $SU(N)$ lattice gauge theories
\vspace*{0.25in}}

\author{Barak Bringoltz and Stephen R. Sharpe\\
\vspace*{0.25in}}

\affiliation{Physics Department, University of Washington, Seattle, WA 98195-1560, USA
\vspace*{0.6in}
}

\begin{abstract}

We study the validity of the large-$N$ equivalence between
four-dimensional $SU(N)$ lattice gauge theory and its momentum
quenched version---the Quenched Eguchi-Kawai (QEK) model. 
We find that the assumptions needed for the proofs of equivalence
do not automatically follow from the quenching prescription. 
We use weak-coupling arguments to show that large-$N$ equivalence is in fact likely to break down in the QEK model, and that this is due to 
dynamically generated correlations between different Euclidean components of the gauge fields.
We then use Monte-Carlo simulations at intermediate couplings
with $20\le N\le 200$ to provide strong evidence for
the presence of these correlations and for the consequent breakdown
of reduction.
This evidence includes a large discrepancy between the transition
coupling of the ``bulk'' transition in lattice gauge theories and the
coupling at which the QEK model goes through a strongly first-order
transition. To accurately measure this discrepancy we adapt the
recently introduced Wang-Landau algorithm to gauge theories.

\end{abstract}

\maketitle

\section{Introduction}
\label{intro}

QCD simplifies in the `t Hooft limit of a large number of colors,
yet differs from the physical case, $N=3$, by only $O(1/N)$ 
corrections~\cite{tHooft,Witten}.
As a result, it has been a long-standing goal to understand
the properties of QCD at large $N$. This goal has
become even more interesting as string theory
developments, based on gauge/gravity duality, attempt to move
towards predictions for QCD-like theories at infinite $N$
\cite{Zamaklar-review}.
Considerable progress towards this goal has been made 
using numerical lattice computations (as reviewed, for example,
in Refs.~\cite{Mike-dublin-05,Narayanan-Regensburg-07}).
For example, using conventional large volume simulations,
and extrapolating from $N=3-16$, precision results for
several quantities in the large-$N$ limit have been obtained.
In addition, many quantities have been found to depend only
weakly on $N$.

In this paper we reconsider an alternative to conventional large volume
simulations, namely the use of large-$N$ volume reduction, in which 
the infinite space-time volume is collapsed to a single point, with
space-time degrees of freedom repackaged into the $O(N^2)$ color
degrees of freedom. This allows one, in principle, to trade two large
parameters, the volume and $N^2$, for a single large parameter,
and consequently to consider much large values of $N$.

The idea of large-$N$ reduction for lattice gauge theories was first
proposed by Eguchi and Kawai in Ref.~\cite{EK}. They defined the
``reduced'' theory to be a lattice gauge theory on a single site and
showed that, under certain assumptions, Wilson loops in the reduced
theory and in the $d$-dimensional infinite-lattice gauge theory acquire the same
expectation values in the large-$N$ limit.  The two main
assumptions in \cite{EK} are that expectation values of products of
single-trace operators factorize at large-$N$ 
(see \Eq{condition1} below)
and that the vacuum is symmetric under the $(Z_N)^d$ center
transformations applied to the model's $SU(N)$
 ``link'' matrices $U_\mu$ (see \Eq{refl_sym})\footnote{
We consider the $SU(N)$, rather than the $U(N)$, theory in this paper.
The two theories become equivalent as $N\to\infty$, but the extra
phase in the $U(N)$ theory, which decouples from the dynamics at any $N$,
obscures the underlying mechanisms we discuss in later sections.}. It was quickly realized, based
on weak coupling arguments and numerical results, that the second
assumption does not hold for $d>2$ in the continuum~\cite{BHN1,MK,Okawa1}, and
various ideas for solving this problem were suggested. The first,
which we focus on here, was of quenching the eigenvalues
of the link matrices by forcing them to have a $Z_N$ invariant
distribution. This means that, by construction, all order parameters
for the breakdown of the $(Z_N)^d$ symmetry, such as $\<\tr U_\mu\>$,
vanish.

This ``quenched Eguchi-Kawai'' (QEK) model was proposed in
Ref.~\cite{BHN1} (see also \cite{Migdal}), and first numerical results
indicated that it did indeed solve the problem of unwanted symmetry
breaking~\cite{BHN2,Okawa2,Bhanot,BM,Lewis,Carlson}.  Intensive
analytic investigation of the QEK model ensued. (For example, see the
papers \cite{HN,GK,DW,Parisi-papers,Neuberger-permutations,Parsons}, which are relevant for us
here.) For further discussion and references we refer to the reviews in
Refs.~\cite{Das-review,Makeenko-book,Migdal-review}.

Studies of the QEK model tailed off, partly because of the emergence
of the apparently more promising ``twisted Eguchi Kawai'' (TEK) model
\cite{TEK}. We do not discuss this model here, but note only that
despite the early literature, recent extensive simulations find evidence that large-$N$
reduction fails in the TEK model~\cite{TV,Ishikawa,Bietenholz}.

Another line of development, initiated in Ref.~\cite{KNN} and extended
in a series of papers summarized in \cite{Narayanan-Regensburg-07},
has been much more successful. This is the idea of partial
reduction. Here one reduces not to a single-site but to a lattice of
size $L^d$, with $L> L_{\rm min}\approx 1\,$fm. As long as the
dimensions exceed this minimum, the $(Z_N)^d$ center symmetry is seen
to be unbroken, so reduction holds and one obtains volume-independent
results if $N$ is sufficiently large. Detailed studies have
demonstrated partial reduction, and provided results for gauge and
fermionic quantities for $N\stackrel{<}{_\sim} 50$~\cite{Narayanan-Regensburg-07}.

In this paper we return to the original QEK model, and study whether
reduction holds there as well, as was indicated by the early works
\cite{BHN2,Okawa2,Bhanot,BM,Lewis,Carlson}. We are motivated to do so
not just as a tool to study very high values of $N$, but also for the
following reason. It has recently been argued
that the $(Z_N)^d$ breaking does not occur in
any volume if the fermions are in the adjoint representation of the
gauge group, and that in such a theory, complete large-$N$ reduction
to a single site should hold~\cite{QCDadjoint}. 
This theory is of phenomenological
interest because, through the orientifold large-$N$ equivalence of
adjoint and two-index antisymmetric tensor fermions, it differs from
physical QCD by corrections of $O(1/N)$ \cite{ASV}. Studying this
theory numerically at any volume, however, is expensive since it
requires simulating dynamical fermions.\footnote{%
In contrast to the `t
Hooft limit where the fermions are in the fundamental representation
and affect the dynamics at $O(1/N)$, in this theory they enter at
$O(1)$ and thus dynamical simulations are necessary.}
Thus, before
plunging in to such a study, we chose to get experience with
single-site models, and the QEK model is a natural candidate.

The outline of paper is as follows. In Section~\ref{EK-review} we
review the features of the original Eguchi-Kawai (EK) model which are
relevant for our study, such as the failure of reduction in this
model, and its relation to $(Z_N)^d$ breaking. In
Section~\ref{QEK-review} we discuss the QEK model in some
detail. We begin by introducing the quenching prescription, and by
presenting the symmetries of the model. We then review the theoretical
arguments for the validity of the large-$N$ equivalence between the
QEK and large-$N$ QCD. In Section~\ref{analytics} we present
weak-coupling analytic considerations that question these arguments
and we conclude that, similarly to the EK model, it is likely that
reduction fails in the continuum limit. In Section~\ref{numerics} we
present an extensive numerical lattice study of the QEK and find
evidence to support our claims of Section~\ref{analytics}. We conclude
in Section~\ref{summary} with remarks on the implications of our
findings for other large-$N$ reduced models, such as those in \cite{DEK}.

\section{Review of the original Eguchi-Kawai model}
\label{EK-review}

We begin with a brief review of the original EK
model, so as to set notation and provide background for our
observations. The EK model is a $d$-dimensional
lattice gauge theory restricted to a
single site whose partition function is
\begin{eqnarray}
Z_{EK} &=& \int DU \, \exp({S_{EK}})\,, 
\label{Z_EK} 
\\ S_{EK} &=& N b
\sum_{\mu<\nu} 2 {\rm Re}\,\Tr(U_\mu U_\nu U_\mu^\dagger
U_\nu^\dagger)
\,.
\label{eq:SEK}
\end{eqnarray}
Here $b=1/\lambda$, with $\lambda=g^2 N$ the 't Hooft coupling, and
the integration measure is the Haar measure on $SU(N)$.  Aside from
the ``reduced'' gauge symmetry,
\begin{equation}
\forall \mu\,: \quad U_\mu \to \Omega \, U_\mu \, \Omega^\dag \qquad 
{\rm with}\qquad \Omega \in SU(N), \label{gauge_sym}
\end{equation}
the model is also symmetric under center transformations
applied independently to the $d$ link matrices
\begin{equation}
U_\mu \to U_\mu \, z^{n_\mu} \quad {\rm with} \quad z = e^{2\pi i N} \quad {\rm and}\quad n_{\mu} \in Z_N \,, 
\label{ZN_sym}
\end{equation}
and under the $d$ reflections 
\begin{equation}
{\cal P}_\mu\,:\qquad U_\mu \to U^\dag_\mu \,. 
\label{refl_sym}
\end{equation}

A Wilson loop on the original lattice that is defined by the path
$C:(x,x+\hat \mu,\dots,x-\hat \nu-\hat \rho,x-\hat \nu,x)$, 
and that is given by
\begin{equation}
W_C = \frac1N \, \tr \, U_{x,\hat \mu} U_{x+\hat \mu,\hat \nu} \cdots
U_{x-\hat \nu-\hat \rho,\hat \rho} U_{x-\hat \nu,\hat \nu}\,,
\end{equation}
is mapped in the reduced model to
\begin{equation}
W^{\rm reduced}_C = \frac1N \, \tr \, U_{\mu} U_{\nu} \cdots U_{\rho}
U_{\nu}\,. \label{open_loop}
\end{equation}
The essence of reduction is that the expectation values of $W_C$
in the gauge theory and of $W^{\rm reduced}_C$ in the reduced model are the same at large-$N$:\footnote{See, however, the discussion of connected correlation functions in \cite{QCDadjoint}.}
\begin{equation}
\<W_C\>_{\rm gauge\,\,theory}=\<W^{\rm reduced}_C\>_{\rm reduced} + O(1/N^2). \label{equivalence}
\end{equation}
To obtain \Eq{equivalence} one derives the Dyson-Schwinger equations
of $W_C$ and $W^{\rm reduced}_C$ and finds that they are the same, up
to additional source terms. These source terms vanish provided that the
following two conditions are satisfied
\begin{eqnarray}
\<W_{C_1}\, W_{C_2}\>_{\rm reduced} 
&=& 
\<W_{C_1}\>_{\rm reduced}\, \< W_{C_2}\>_{\rm reduced} + O(1/N^2),\label{condition1} \\
\<W_{\rm open}\>_{\rm reduced} &=& 0.
\label{condition2}
\end{eqnarray}
Here \Eq{condition1} must hold  for all contours $C_1$ and $C_2$,
while in \Eq{condition2} $W_{\rm open}$ denotes any reduced Wilson loop
whose path is a mapping of an open path in the gauge theory, e.g.
\begin{equation}
C_{\rm open}: (x,x+\hat \mu,\dots,y-\hat \nu,y) \quad ; \quad y\neq x\,.
\end{equation}
This means that, in at least one direction, the corresponding
$W_{\rm open}^{\rm reduced}$ of \Eq{open_loop} has the
the number of $U$'s less the number of $U^\dag$'s different from zero.\footnote{%
More precisely, this difference need only be zero modulo $N$, a subtlety
that will not play any role in our considerations.}

The factorization required in \Eq{condition1} is expected to hold
in large-$N$ theories, 
and realizes the idea of the ``Master field''~\cite{Witten}. 
Condition (\ref{condition2}) holds if the
vacuum is $(Z_N)^d$ symmetric, since under this symmetry $W_{\rm open}$
acquires a phase.
In fact, as already noted in
Section~\ref{intro}, this symmetry is spontaneously broken in the
four-dimensional EK model, at weak-coupling. This is shown
by perturbative calculations in the weak-coupling
($b\to \infty$) limit~\cite{BHN1,MK,Zeromodes}, and has been demonstrated
by lattice simulations to hold for $b\stackrel{>}{_\sim}0.19-0.30$~\cite{BHN1,Okawa1}.
Thus \Eq{condition2} does not hold, invalidating reduction.

It is useful for our subsequent discussion to briefly
recall the perturbative calculation of Refs.~\cite{BHN1,MK}. 
One first writes the link matrices in polar form
\begin{equation}
U_\mu = V_\mu^\dagger \Lambda_\mu V_\mu\,,
\label{eq:formofUmu}
\end{equation}
where $V_\mu\in SU(N)$ and $\Lambda_\mu$ is a diagonal matrix
containing the eigenvalues,
\begin{equation}
\Lambda_\mu = {\rm diag}[\exp(ip_\mu^1),\dots,\exp(ip_\mu^N)]\,; 
\qquad p^a_\mu \in [0,2\pi)\,.
\label{eq:momentadef}
\end{equation}
For $U_\mu\in SU(N)$, the $p_\mu^a$ are constrained to satisfy\footnote{%
We use $a$ as a color index, and denote the lattice spacing by
$a_{\rm lat}$.}
\begin{equation}
\sum_{a=1}^N p_\mu^a = 0\ {\rm mod}\ N 
\label{eq:SUNconstraints}
\end{equation}
in each direction---a constraint that we keep implicit
in the following formulae for the sake of clarity. 
Using \Eq{eq:formofUmu} one can show that the
partition function \Eq{Z_EK} becomes
\begin{equation}
Z_{EK} = \int \prod_{\mu,a} \frac{dp^a_\mu}{2\pi} \, \Delta^2(p) \int
DV \exp{S_{EK}}
\equiv \int \prod_{\mu,a} \frac{dp^a_\mu}{2\pi} \, \exp
{\left[-F_{EK}(p)\right]}
\,,\label{Z_EK2}
\end{equation}
where
\begin{equation}
\Delta^2(p) \equiv \prod_\mu \prod_{a<b} 
\sin^2 \left( \frac{p^a_\mu-p^b_\mu}{2}\right) \label{VdM}
\end{equation}
is the Vandermonde determinant, and $DV$ is the Haar measure on
$SU(N)$. 
For all values of $p^a_\mu$, 
the action $S_{EK}$ is minimized when $V_\mu=1$ for all $\mu$
(up to gauge transformations).
Expanding $V_\mu$ around unity, 
and assuming non-degenerate $p$, i.e. $p^a_\mu\neq p^b_\mu$ if $a\neq b$, 
one finds at leading order
in the weak-coupling expansion that the free energy is~\cite{BHN1,MK}
\begin{equation}
F_{\rm EK}(p) \stackrel{b\to\infty}{\longrightarrow} (d-2)\sum_{a<b} \log\left[
\sum_\mu \sin^2\left(\frac{p^a_\mu-p^b_\mu}2\right) \right].
\label{eq:freeenergy}
\end{equation}
For $d>2$, $F_{\rm EK}(p)$
is minimized when, for each $\mu$, the $p^a_\mu$ for all $a$ become equal.\footnote{%
As noted in Ref.~\cite{BHN1},
the weak-coupling calculation of $F_{\rm EK}$ breaks down
for the degenerate eigenvalues that are picked out
by minimizing $F_{\rm EK}$. 
This is due to the presence of zero modes.
One can extend the calculation to include the effects
of these modes and the conclusions are unchanged
in the weak-coupling limit~\cite{Zeromodes}.
At moderate values of $b$, however, only numerical calculations are reliable, 
and these~\cite{BHN2,Okawa1} are consistent with 
the weak-coupling picture of \cite{BHN1}.}
This implies that, for $SU(N)$, the theory has $N^d$ ``vacua'',
with $U_\mu\approx e^{ip_\mu^a} {\bf 1} \approx z^{m_\mu} {\bf 1}$,
which are transformed into each other by
the center transformations (\ref{ZN_sym}).

Such a clustering of the eigenvalues appears to
indicate spontaneous breakdown of the center symmetry and so to invalidate \Eq{condition2}.
To establish spontaneous symmetry breaking (SSB), however, 
one needs to know whether fluctuations
about each  vacuum are sufficient to restore
the symmetry (as happens for $d\le 2$ in infinite volume statistical
mechanical systems). In other words, does the free-energy barrier
between the different vacua become infinitely high as $N\to\infty$
(implying symmetry breaking) or not (implying symmetry restoration)?
Also, the calculation leading to \Eq{eq:freeenergy} is valid only when 
$b\to \infty$, and it is possible that the symmetry is restored for 
moderate values of $b$. In fact, as noted above, numerical simulations imply 
that the symmetry is indeed broken once $b$ becomes moderately large.

We conclude this section with a general remark
concerning the possible ways that reduction can fail.
Of the two key conditions, Eqs.~(\ref{condition1})-(\ref{condition2}), 
it is often considered to be the second that is crucial.
Thus the validity of Eguchi-Kawai reduction has become 
almost synonymous with the 
absence of spontaneous breaking of the center symmetry. 
In this scenario there are multiple vacua, connected by
symmetry transformations, around each of which
the fluctuations are $\sim 1/N^2$.
We wish to emphasize, however, that this is not the only possibility. 
What is required for reduction to hold is 
the {\em combination} of \Eq{condition1} and \Eq{condition2},
and it is also possible that the first of these can fail.
This happens if there are multiple would-be
symmetry-breaking ground states yet fluctuations lead to
motion between all these states even when $N\to\infty$. 
(This possibility
has been already mentioned in the previous paragraph.)
In an infinite volume theory this corresponds to a breakdown
of cluster decomposition.
These two scenarios should be compared to that
with a {\em single} vacuum obeying cluster decomposition,
in which case both relations hold and reduction is valid.

The distinction between the failure of reduction due to
a breakdown of cluster decomposition and due to SSB
is important below, so we illustrate it with
the following simple example.
For an $SU(N)$ EK theory, as noted above, there is SSB for sufficiently
weak coupling, with $N^{d}$ vacua.
If we change the gauge group to $U(N)$, however, 
the free energy landscape has a ``Mexican-hat''
form, and the $N^d$ vacua become part of a continuous
degenerate manifold connected by the four $U(1)$ phases (one per
direction). The path integral over these phases causes
expectation values of open loops to vanish and \Eq{condition2} holds.
This does not, however, mean that reduction holds, because the first
of the required conditions, eq.~(\ref{condition1}), is not satisfied.
The failure of (\ref{condition1}) can be seen by comparing its two
sides for the case where $C_1$ and $C_2$ are both open loops that,
when combined, form a closed loops (for example see Fig.~\ref{W-loop}). For
this choice the $U(1)$ transformations multiply $W_{C_1}$ and
$W_{C_2}$ by opposite phases and so the l.h.s. of \Eq{condition1} is
independent of these $U(1)$ phases and has an $O(1)$ value. In
contrast, the r.h.s. is of $O(1/N^2)$ because the open-loop
expectation values do vanish. The failure of factorization is perhaps
surprising, but occurs because the vacuum is not unique and because
the $U(1)$ degrees of freedom have unconstrained fluctuations. This is
also an example of why it is simpler to analyze the $SU(N)$ theory.

\section{Review of the quenched Eguchi-Kawai model}
\label{QEK-review}

Quenching attempts to avoid the spontaneous breaking of the $(Z_N)^d$
symmetry by forcing the eigenvalues not to cluster. In this section we
recall the quenching prescription and describe its relation to infinite-volume
large-$N$ QCD.

\subsection{The quenching prescription}
\label{QEK_pres}

The prescription consists of
first calculating expectation values for a fixed
set of the eigenvalues $p^a_\mu$ (labeled collectively by ``$p$'' and
distinguished from the usual expectation values by a subscript),
\begin{equation}
\left\langle{{\cal O(U)}}\right\rangle_p \equiv Z(p)^{-1}
\int \prod_\mu DV_\mu \,\,e^{S_{\QEK}(p)} \,\, {\cal O(U)}
\,.
\label{eq:O(U)theta}
\end{equation}
Here $S_\QEK(p)$ is simply $S_{\rm EK}$ with the $U_\mu$ having the
form (\ref{eq:formofUmu}), i.e.
\begin{equation}
S_{\rm QEK}(p) = N b \sum_{\mu<\nu} 2 {\rm Re}\,\Tr\left((V_\mu^\dagger
\Lambda_\mu V_\mu)\, (V_\nu^\dagger \Lambda_\nu V_\nu )\,
(V_\mu^\dagger \Lambda_\mu^\dagger V_\mu )\, (V_\nu^\dagger
\Lambda_\nu^\dagger V_\nu)\,\right)
\,,
\label{eq:SQEK}
\end{equation}
$Z(p)$ is the partition function for fixed $p$,
\begin{equation}
Z(p) \equiv \int \prod_\mu DV_\mu \exp(S_{\rm QEK}(p))\,,
\label{eq:ZQEK}
\end{equation}
and the $V_\mu$ take values in $SU(N)$. 
The second part of the prescription is
to average expectation values over the choices of $p$:
\begin{equation}
\left\< {\cal O(U)} \right\>_\QEK = \int dp \, \,
\left\langle{{\cO(U)}}\right\rangle_p \equiv 
\int_0^{2\pi} \prod_{\mu,a} \frac{dp_\mu^a}{2\pi} \;
\rho(p) \, 
\left\langle{{\cO(U)}}\right\rangle_p 
\,.
\label{eq:QEKexpectationvalue}
\end{equation}
Here $\rho(p)$ is a positive weight function (with integral normalized to
unity) which is invariant under the $(Z_N)^d$ shifts $p^a_\mu \to p^a_\mu + 2\pi k_\mu/N$,
and dense in the space of the $p_\mu^a$ as $N\to\infty$.
For the $SU(N)$ theory, it should also incorporate the 
constraints (\ref{eq:SUNconstraints}).
We discuss particular choices of $\rho(p)$ below.

To understand the significance of quenching, it
is useful to write expectation values in the {\em original} EK model
in terms of the $p$-dependent $\<{\cal O(U)}\>_p$:
\begin{equation}
\left\< {\cal O(U)} \right\>_{\rm EK} = 
\frac{
\int_0^{2\pi} \, \prod_{\mu,a}  \frac{dp_\mu^a}{2\pi} \,\, \Delta^2(p)
\, Z(p) \, \left\langle{{\cO(U)}}\right\rangle_p }
{\int_0^{2\pi} \, \prod_{\mu,a} \, \frac{dp_\mu^a}{2\pi} \,
\Delta^2(p) \, Z(p)}
\,.
\label{eq:EKexpectationvalue}
\end{equation}
Comparing \Eq{eq:EKexpectationvalue} and \Eq{eq:QEKexpectationvalue}
 we see that quenching changes the measure of the integral over 
the $p$ in such a way as to replace the non-uniform
weighting $\Delta^2(p) Z(p)=\exp[-F_{\rm EK}(p)]$, which was the cause of
the clustering of eigenvalues~\cite{Migdal}, 
with the uniform weighting $\rho(p)$.

\subsection{Symmetries in the quenched model}
\label{QEK_sym}

Since quenching has divided the original dynamical degrees of
freedom (the $U_\mu$) into the dynamical $V_\mu$ and the quenched $p_\mu^a$,
it is important to understand how the symmetries 
Eqs.~(\ref{gauge_sym})-(\ref{refl_sym}) are realized.
The gauge transformations \Eq{gauge_sym} can be chosen to act
only on the $V_\mu$:
\begin{equation}
V_\mu \to V_\mu \Omega \,. \label{gauge_sym_QEK}
\end{equation}
By contrast, the center and reflection symmetries must,
in general, be realized by transforming the eigenvalues:
center transformations (\ref{ZN_sym}) become
\begin{equation}
p_\mu^a \to {\rm mod} \left(p_\mu^a + \frac{2\pi n_\mu}{N},2\pi\right)
\,,
\label{eq:momentaZN}
\end{equation}
while the reflections (\ref{refl_sym}) become
\begin{equation}
p_\mu^a \to 2 \pi - p_\mu^a
\,.
\label{eq:momentarefl}
\end{equation}
There are, however, special choices of $\rho(p)$ for which one can
realize Eqs.(~\ref{ZN_sym}-\ref{refl_sym}) by transformations on the
$V_\mu$ alone, and we discuss these in Section~\ref{QEK_equiv} below.

Quenching solves the problem of the dynamical clustering of
eigenvalues, and leads to the desired vanishing of the expectation
values of open Wilson loops such as $\tr(U_\mu)$. To see this, note
that the center transformation \Eq{eq:momentaZN} performs a ``clock
rotation'' of the $p_\mu^a$ and thus multiplies $\left\langle
\tr(U_\mu)\right\rangle_p$ by $z^{n_\mu}$. This is as in the unquenched EK
model, but now the $p_\mu^a$ are forced to have a $Z^d_N$ symmetric distribution. Thus the integration in \Eq{eq:QEKexpectationvalue} leads to
\begin{equation}
\left\langle  {\tr(U_\mu)}\right\rangle_\QEK 
= \int dp \,\, \left\langle\tr (U_\mu)\right\rangle_p\,\,= 0\,.
\label{eq:trUQEK}
\end{equation}
In fact, the quenched expectation value of any open loop will vanish
due to the (now enforced) average over the center transformations.

\subsection{Large-$N$ reduction in the QEK model}
\label{QEK_equiv}

\begin{figure}[bt]
\centerline{
\includegraphics[width=3cm]{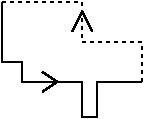}
}
\caption{Illustration of the extra source terms,
eq.~(\protect\ref{eq:loop0}), in the
Dyson-Schwinger equations for a closed Wilson loop in the QEK model.
The dashed line represents $W_{\rm open}$ while the solid line
represents $W'_{\rm open}$.}
\label{W-loop}
\end{figure}

As in the EK model, one can derive the Dyson-Schwinger equations for
Wilson loops in the QEK model. They too include unwanted
source terms. In the QEK model, some of
these have the form~\cite{GK,Parisi-papers}\,\footnote{Here we note that the derivation of the Dyson-Schwinger equations in the QEK model is different than in the EK model, and in addition to the terms  of the form of  \Eq{eq:loop0}, there are other source terms which have a similar, but more complicated form \cite{GK}. Nevertheless, the analysis we perform in this section holds for these terms as well.},
\begin{equation}
\left\langle W_{\rm open}
W'_{{\rm open}}\, \right\rangle_\QEK = \int dp\,\,
\left\langle W_{\rm open} \,\, W'_{{\rm open}}\right\rangle_p\,
\label{eq:loop0}
\end{equation}
where $W_{\rm open}$ and $W'_{{\rm open}}$ are open Wilson loops 
(as defined in Section~\ref{EK-review}) that, when joined, form
a closed loop, as illustrated in Fig.~\ref{W-loop}.
Such terms must all vanish for reduction to hold.  

The argument that
they do vanish proceeds in two steps~\cite{HN,GK,Parisi-papers}:
\begin{eqnarray}
\int dp\,\,  \left\langle W_{\rm open} \,\,
W'_{{\rm open}}\right\rangle_p\, 
&=&
\int dp\,\, \left\langle W_{\rm open} \right\rangle_p\,\,
\left\langle W'_{{\rm open}}\right\rangle_p 
+\, O(1/N^2) \label{eq:loop1}
\\
\int dp\,\, \left\langle W_{\rm open} \right\rangle_p\,\,
\left\langle W'_{{\rm open}}\right\rangle_p 
&=&
\int dp\,\,  \left\langle W_{\rm open} \right\rangle_p\
\int dp' \left\langle W'_{{\rm open}}\right\rangle_{p'}
+\, O(1/N)\,.
\label{eq:loop2}
\end{eqnarray}
The first step is large-$N$ factorization for a fixed set of $p$,
valid to all orders in perturbation theory.  The second step, which
one might call ``quenched factorization'', is special to the quenched
theory.  If it holds, then, due to the vanishing of quenched
expectation values of open loops [as in \Eq{eq:trUQEK}] the extra terms 
(\ref{eq:loop0}) in the loop equations do vanish in the large-$N$ limit.

We will argue in subsequent sections that the combination of
eqs.~(\ref{eq:loop1}) and (\ref{eq:loop2}) does not hold,
most likely due to a failure of the latter equation.
In order to understand what fails, we must first describe the argument for the
correctness of these steps, and indeed why the quenching prescription is expected to reproduce the large-$N$ dynamics of $SU(N)$ lattice gauge theories.

\bigskip
The approach of Refs.~\cite{BHN1,HN,GK,Parisi-papers,DW} provides an intuitive 
explanation of why large-$N$ quenched reduction works for a wide class
of theories, albeit within perturbation theory. The idea relies on the
fact that, in large-$N$ perturbation theory, only planar diagrams
survive. In `t Hooft's double-line notation each gluon line is replaced
by two oppositely pointing lines that carry two indices $(a,b)$ with
$a,b\in [1,N]$.  In the reduced theory there is, initially, no momentum
associated with a gluon ``propagator''.
The key point is then to associate a $d$-dimensional
lattice momentum with each of the indices,
\begin{equation}
a \ \ \leftrightarrow \ \ p^a_\mu \,; 
\qquad  p^a_\mu \in [0,2\pi)\,, \label{mom-index}
\end{equation}
and to assign to the gluon with color indices $(a,b)$ the
difference in the momenta associated with the two indices:
\begin{equation}
q^{ab}_\mu\equiv {\rm mod} \left( p^a_\mu-p^b_\mu \, ,2\pi\right)\, \label{q_diff_p}\,.
\end{equation}
It is easy to check
 that since one lets all $p^a_\mu$ take all values in $[0,2\pi)$, 
 the momenta $q_\mu^{ab}$ of all gluons in any planar diagram will
 independently take values in the Brillouin zone, and will obey momentum
 conservation (modulo $2\pi$) at the vertices. This is impossible for
 non-planar diagrams, where some of the gluons carry the indices
 $(a,a)$ and so have $q^{aa}_\mu=0$.

The identification of \Eq{mom-index} is necessary in order to embed
space-time (or rather the first Brilluoin Zone of its momentum space) 
in color space, but
it is only the first step. The next is to choose the
action of the single-site model in such a way that the actual
value of planar diagrams will be the same as that in the full gauge
theory. In particular, they choose it such that the $(a,b)$ matrix
element of the gluon propagator takes (in an appropriate gauge) the
usual ``$1/q^2$'' form (or, more precisely, its lattice version) with
$q$ indeed being the {\em difference} $p^a-p^b$.  Vertices are
similarly reproduced.  In this way, for a given choice of color
indices, one obtains the correct {\em integrand} of the corresponding
infinite-volume large-$N$ Feynman diagram.

The application of the quenching prescription 
involves an important subtlety, which requires
that additional constraints be placed on the single-site
fields~\cite{GK,DW}.  The end result is that one arrives at precisely the
QEK model, with the quenching prescription of Eqs.~(\ref{eq:O(U)theta})--(\ref{eq:QEKexpectationvalue}).  The quantities $p_\mu^a$
are now viewed as (dimensionless) loop momenta. For example,
perturbing around the classical vacuum of $V_\mu=\bm{1}$, yields the
following two-point function for the fluctuating matrices $A^{ab}_\mu$
\begin{equation}
\langle A_\mu^{ab} A_\nu^{b'a'} \rangle_p
= \delta_{\mu\nu}\delta_{aa'}\delta_{bb'}
\frac{4}{\sum_\nu \sin^2\left(\frac{p_\nu^a-p_\nu^b}{2}\right)}
\qquad (a\ne b), \label{gluon_prop}
\end{equation}
which is the standard lattice result for the gluon
propagator.\footnote{%
As explained in \cite{GK,DW}, the expansion used to
obtain \Eq{gluon_prop} is $V_\mu \Lambda_\mu V^\dag_\mu
\Lambda^\dag_\mu = e^{ig A_\mu}$.}

The final step in the quenching prescription of Ref.~\cite{BHN1,GK,Parisi-papers,DW} is to
average over the momenta, as in
eq.~(\ref{eq:QEKexpectationvalue}). The 
weight function 
should be manifestly $(Z_N)^d$ invariant, and,
as $N\to\infty$, force the
momentum-components in each direction to densely cover the Brilluoin zone. 
An example for such a measure is that  suggested  in~\cite{BHN2} : 
\begin{equation}
\rho_{\rm VdM}(p) = \Delta^2(p)\,,
\label{eq:Weylweight}
\end{equation}
where the Vandermonde determinant is defined in \Eq{VdM}. 
At large-$N$ the function $\rho_{\rm VdM}(p)$ forces the momenta 
to lie as far apart from each other as possible, and combined with the $SU(N)$ constraints
(\ref{eq:SUNconstraints}), this requires the $p_\mu^a$ to be a
permutation of the ``clock'' values,%
\footnote{Note that here the Brilluoin zone is, at large-$N$, 
$[-\pi,\pi)^d$, instead of $[0,2\pi)^d$, but this change is irrelevant in our discussion.}
\begin{equation}
P^a = \frac{2\pi}{N} \left(a- \frac{N+1}{2}\right)\,,\qquad
a\in[1,N]\,.
\label{eq:clockmomenta}
\end{equation}
The integral over $p$ then amounts to an average over permutations,
independently for each direction.  Since the momenta
(\ref{eq:clockmomenta}) are uniformly distributed, this gives a
discrete approximation to the integration \Eq{eq:QEKexpectationvalue}
over the infinite-volume momentum space.  Thus as $N\to\infty$ one
obtains, order by order in perturbation theory, the correct
infinite-volume result for each Feynman diagram.

We can now give the argument of Refs.~\cite{GK,Parisi-papers} for the crucial
relation (\ref{eq:loop2}).  Imagine evaluating the l.h.s. of \Eq{eq:loop2} 
in perturbation theory, and focus on the contribution from planar diagrams 
with $(L+M)$-gluon loops, with $L$ loops coming from the expansion of $W_{\rm open}$, 
and the remaining $M$ loops from the expansion of $W'_{\rm open}$. 
We can write this contribution as
\begin{equation}
\int dp \, \sum_{a_1,a_2,\dots,a_{L} \atop b_1,b_2,\dots,b_{M} } f(p_{a_1},p_{a_2},\dots,p_{a_L})\, g(p_{b_1},p_{b_2},\dots,p_{b_M}). \label{GK_argument1}
\end{equation}
Here $f$ and $g$ denote the integrands of the planar diagrams
contributing to the two Wilson loops, and $p_{a_i}$ and $p_{b_i}$ are
the momenta that flow in these diagrams.  If $(L+M)\ll N$ then a
generic term in the double sum has all indices different, and for each
such term the (normalized) integral over $p$ factorizes in the large-$N$ limit into the two
integrals
\begin{equation}
\int dp \, f(p_{a_1},p_{a_2},\dots,p_{a_L})\, g(p_{b_1},p_{b_2},\dots,
p_{b_M}) = \int dp \, f(p_{a_1},p_{a_2},\dots,p_{a_L})\, 
\int dq \, g(q_{b_1},q_{b_2},\dots,q_{b_M})\,. 
\label{GK_argument2}
\end{equation}
Here, for clarity, we changed the dummy integration variables
$p_{b_i}$ to $q_{b_i}$. The last step is to sum
\Eq{GK_argument2} over {\em all possible values} of the indices
$a_i$ and $b_i$, which gives the $L$-loop contribution to $\<W_{\rm
open}\>$ multiplied by the $M$-loop contribution to $\<W'_{\rm
open}\>$. If this step is correct, we obtain the r.h.s. of
\Eq{eq:loop2}.

This last step is only approximately correct for the following reason:
\Eq{GK_argument2} holds only if the indices $a_{1,\dots,L}$ are all
different from the indices $b_{1,\dots,M}$, while in performing the
final sum we ignore this restriction. At large-$N$, however, the
effect of this ``negligence'' is only of $O(1/N)$---the fraction of
terms with equal $a_i$ and $b_i$ indices. As a result one finds that
to all orders in planar perturbation theory, quenched reduction holds
at large-$N$. The fact that one ignores $O(1/N)$ terms here also
explains why the quenched model has $O(1/N)$  rather than
$O(1/N^2)$ corrections.

\subsection{Alternative choices of $\rho(p)$}
\label{QEK_rho_choices}

The formulation of the QEK model provided by the
approach of Refs.~\cite{GK,Parisi-papers} shows that there is
considerable freedom in choosing the weight function $\rho(p)$. 
The choice must simply turn the integrands of Feynman diagrams into their integrals
as $N\to\infty$.
The simplest choice is to
use a uniform distribution: $\rho_{\rm uniform}(p)=1$ for all $N$. In practice, this
can be implemented by Monte-Carlo---drawing $p$ randomly from a
uniform distribution.  Another choice, which we call $\rho_{\rm clock}(p)$,
is to take the momenta to be a permutation of the clock
values of eq.~(\ref{eq:clockmomenta}), even for finite $N$, and then
average over permutations. 
This corresponds to setting
\begin{equation}
\rho_{\rm clock}(p) = \frac{1}{(N!)^4}\left(\prod_\nu \sum_{\sigma_\nu}\right) 
\left\{\prod_{\mu,a} \delta(p_\mu^a-P^{\sigma_\mu(a)})\right\}\,,
\label{eq:rho_clock_def}
\end{equation}
where the $\sigma_\mu(a)$ are permutations of the color indices.
With this choice, which we use extensively
below, the center-symmetry-breaking order parameters $\tr\,U_\mu^p$
(for $p \ne 0\ {\rm mod}\ N$) vanish {\em prior} to the average over permutations.
This choice also has a simple physical interpretation.  The
discrete $p_\mu$ are those that one would obtain if one had a lattice
with $N$ sites in each direction, with periodic/antiperiodic boundary
conditions for $N$ odd/even.  After averaging over permutations one
obtains the result of the Feynman diagram on a lattice of physical
volume $(Na_{\rm lat})^d$, where $a_{\rm lat}$ is the lattice spacing.

It is further argued in Ref.~\cite{GK,Parisi-papers} that the integral over
momenta (or sum over permutations) becomes unnecessary as
$N\to\infty$.  This is because \Eq{q_diff_p} implies that the sum over
the color indices in each color loop becomes an integral over the
corresponding Brillouin zone. Thus a single choice of randomly chosen
momenta, or a single set of randomly chosen permutations of clock
momenta, should, in principle, be sufficient.  In practice, for finite
$N$, it may be preferable to include an average over such choices.

\begin{figure}[htb]
\centerline{
\includegraphics[angle=-90,width=5cm]{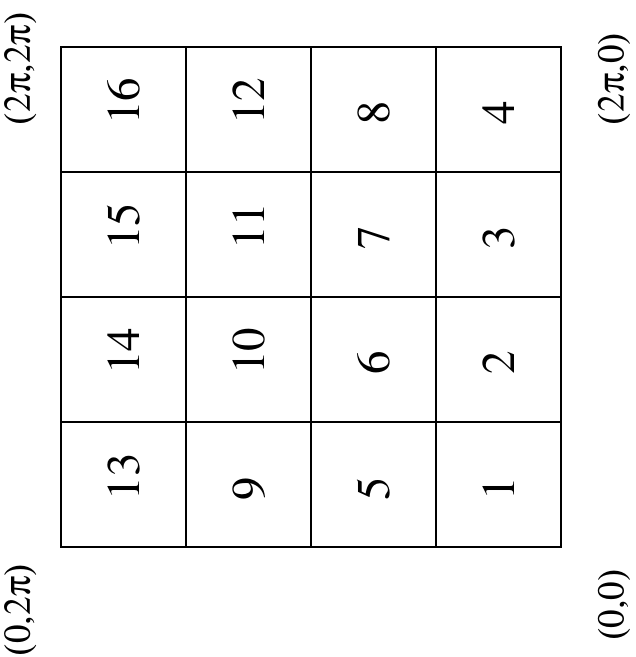}
}
\caption{A two-dimensional example of the embedding suggested in Refs.~\cite{GK,Bars} 
of the color indices in the Brilluoin zone for $N=16$.}
\label{BarsN16d2}
\end{figure}

The final choice of $\rho(p)$ we consider is that suggested in
\cite{GK} and analyzed by Bars in~\cite{Bars}.  
It applies only when $N=K^d$, where $K$ is an integer.
In this case $\rho(p)$ is a
product of delta-functions such that each value of the color index is
associated with a different $d$-dimensional momentum lying on a $K^d$
latticization of the Brillouin zone (BZ).
In four dimensions one has
\begin{eqnarray}
\lefteqn{
\!\!\!\!\left\{p_1^{a},\, p_2^{a},\, p_3^{a},\, p_4^{a}\right\} = }
\nonumber\\
&&\!\!\!\!\!\!\!\frac{2\pi}{K} \left\{{\rm mod}\left(a\!-\!1,K\right), 
{\rm mod}\left(\left[\frac{a\!-\!1}{K}\right],K\right), 
{\rm mod}\left(\left[\frac{a\!-\!1}{K^2}\right],K\right), 
{\rm mod}\left(\left[\frac{a\!-\!1}{K^3}\right],K\right) \right\}
,
\label{Bars_p}
\end{eqnarray}
where $[x]$ indicates the integer part. We shall henceforth denote this
choice by ``BZ''.
As an example, consider the two-dimensional case with $N=16$ (so
$K=4$). We divide the Brillouin zone 
into
sixteen boxes, and set the sixteen momenta $p^a$ to lie at their centers,
as shown in Fig.~\ref{BarsN16d2}.

For $\rho(p)=\rho_{BZ}(p)$, the momenta
are those for a physical volume $(Ka_{\rm lat})^d=N (a_{\rm lat}^d)$,
which is much
smaller than that obtained using the clock momenta.  The
advantage of using $p=p_{\rm BZ}$ is that one obtains a uniform
distribution over the Brillouin zone from a single set of momenta,
already at finite $N$. 

The general discussion of the realization of center and
reflection symmetries given in
Section~\ref{QEK_sym} must be modified for 
$\rho_{\rm clock}$ and $\rho_{\rm BZ}$.
These symmetries can 
now be realized by a transformation on
the $V_\mu$. Consider first the clock momenta.
There is then an $SU(N)$ matrix $S$,
such that, when $V_\mu \to S V_\mu$,
\begin{equation}
U_\mu = V_\mu^\dagger \Lambda_\mu V_\mu
\to V_\mu^\dagger S^\dagger \Lambda_\mu S V_\mu
= V_\mu^\dagger (\Lambda_\mu z_\mu) V_\mu
= U_\mu z_\mu \,.
\end{equation}
This is possible because, first, up to an overall phase, the 
elements of $\Lambda_\mu$
are a permutation of the N'th roots of unity (so that multiplication
by $z_\mu$ corresponds to a permutation of these elements) and second the eigenvalues of $\Lambda_\mu$
can be arbitrarily permuted by conjugation by $SU(N)$ matrices (as will be discussed in detail in the subsequent section).
The reflection transformations also correspond to permutations of the eigenvalues, 
and can be accomplished by different choices of $S$.

The situation is slightly different for $\rho_{\rm BZ}$. Here
the quenched theory only realizes a $(Z_K)^d$ subgroup of the
center symmetry, because only such transformations correspond
to a permutation of the momenta. The reflection transformations are
also realized by permutations.

\bigskip

\section{Breakdown of quenched reduction - analytic considerations }
\label{analytics}

As explained in the previous section, the validity of quenching is
predicated on the momenta being fixed by hand, independently in each
direction, and then integrated over with a suitable weight function.
This is possible in perturbation theory.  When one does a
non-perturbative calculation, however, the values of $p$ are not
completely fixed, and their distribution is thus determined in part by
dynamics. In other words, they are incompletely quenched. We argue in
this section that, at least in the weak coupling limit $b\to\infty$,
the dynamics is likely to choose a ground state in which this
incomplete quenching invalidates reduction in perturbation theory, and
also invalidates the key relations (\ref{eq:loop1}-\ref{eq:loop2}).
If this persists beyond perturbation theory, then reduction fails. Our numerical results, obtained for finite $b$, suggest that this is indeed what occurs.

The key observation is simply stated.  The ``fixed'' momenta can be
dynamically permuted, independently in each direction, by fluctuations
in the $V_\mu$. These innocent-sounding permutations lead, in general, to a
different free energy, and the dynamics chooses the permutation(s)
with the lowest free energy. The $p$'s that one puts in by hand are
different, in general, from those chosen by the dynamics, and the
latter are not uniformly distributed in the Brillouin Zone.
Thus, the sum over
color indices does not lead to a uniform integration over the
Brillouin zone, and the agreement with infinite-volume perturbation
theory fails.

The presence of permutations in the dynamics has long been known, and
was stressed particularly in Refs.~\cite{Neuberger-permutations,Parsons} and \cite{KNN}. To
our knowledge, however, the implications for the validity of reduction
have not previously been noted.

\subsection{``Momentum locking'' at weak coupling}
\label{sec:locking}

Permutations are generated by transpositions, in which $ p_\mu^a
\leftrightarrow p_\mu^b$ for one pair $(ab)$ and one choice of $\mu$.
Transpositions can be accomplished, for example, by multiplying
$V_\mu$ from the left by the $SU(N)$ matrices $V^{(ab)}(\phi)$, which
are the identity aside from
\begin{equation}
\left(V^{(ab)}(\phi)\right)_{(ab)\ {\rm block}}
= e^{i \phi \sigma_1}
\,.
\label{eq:transp_V}
\end{equation}
As $\phi$ runs from $0$ to $\pi/2$, $V_\mu^\dagger \Lambda_\mu V_\mu$
changes to $V_\mu^\dagger \Lambda'_\mu V_\mu$, where $\Lambda'_\mu$
differs from $\Lambda_\mu$ in having the $a$'th and $b$'th momenta
permuted.  Since it is possible to reach any permutation by a sequence
of transpositions, an ergodic simulation will pass through all
possible permutations of the input momenta.
The question then is which
of these permutations has the smallest free energy. Only if they are
equally likely will the quenched model work as desired.

We can calculate the relative free energies in the weak-coupling
limit. First we describe the energy (i.e. action) ``landscape''.  
The minimum energy
states, after appropriate gauge fixing, have $V_\mu= {\bf 1}$ and the
momenta in {\em any} permutation of their input values.  This is
because the plaquette is unity for any choice of diagonal $\Lambda$'s.
If the momenta are non-degenerate, there are in fact $(N!)^{d-1}$
different ``vacua'' (one factor of $N!$ being removed using a gauge
transformation \Eq{gauge_sym} to keep $\Lambda_{\mu=1}$ in its input
order).  These vacua are connected by the $V^{(ab)}(\phi)$, with the
energy barrier (at $\phi=\pi/4$) being~\cite{Neuberger-permutations,Parsons}
\begin{equation}
-\Delta S_{\rm QEK} = 8 N b\, \sin^2(\Delta p_\mu^{ab}/2)
\sum_{\nu\ne\mu} \sin^2(\Delta p_\nu^{ab}/2)\,. 
\label{eq:Parsonsbarrier}
\end{equation}
Here the transposition is being done on $\Lambda_\mu$, and $\Delta
p_\mu^{ab}= p_\mu^a- p_\mu^b$.  
Generically, all the $\Delta p$ are of $O(1)$, and the barrier height
then grows as $N$.  There will always be some $(ab)$ pairs, however,
that have $\Delta p=O(1/N)$, and for these the energy barrier vanishes 
with increasing $N$.  Nevertheless, for fixed $N$, as $b\to\infty$, these
barriers too become infinitely high.

Thus, in the weak-coupling limit, and assuming non-degenerate momenta,
one can treat the system as a collection of independent vacua with
$V_\mu$ fluctuating around unity in each. 
At leading order the free energy in each vacuum is, 
up to an irrelevant constant,~\cite{BHN1,MK}
\begin{eqnarray}
-\ln Z( p) &=& F_{\rm EK}( p) + F_2( p) 
\,,\\
F_{\rm EK}( p) &=& (d-2)\sum_{a<b} \log\left( \sum_\mu \sin^2[\Delta
p_\mu^{ab}/2] \right) \label{F1}
\\
F_2( p) &=& - \ln \Delta^2(p)
\,.
\label{eq:F2}
\end{eqnarray}
This is just a repetition of the result already quoted
in \Eq{eq:freeenergy}, taking into account the difference in 
the definitions of $Z(p)$ and $Z_{\rm EK}(p)$.
Since $\Delta^2(p)$ is
the same for all permutations, it is only $F_{\rm EK}(p)$ which
distinguishes between them.\footnote{%
Note that if one uses the weight function $\rho_{\rm VdM}(p)$ then
$F_2(p)$ is canceled by the Vandermonde determinants in $\rho_{\rm
VdM}$.  If one uses the clock momenta then $\Delta^2(p)$ is a
constant.}
The argument of the logarithm in $F_{\rm EK}( p)$ is just the lattice gluon
propagator, and the overall factor $d-2$ is the number of transverse
gluons.  The key observation is simply that $F_{\rm EK}( p)$ depends on the
permutation of the momenta. To see this qualitatively, note that,
because the logarithm is a concave function, $F_{\rm EK}(p)$ is minimized by
choosing permutations in which there are values of $(ab)$ for which
$\Delta p_\mu^{ab}$ is simultaneously small in all directions. In
other words, one lowers the free energy by aligning, or ``locking'',
the momenta in different directions.  The gain one makes by locking
the small momenta outweighs the loss incurred by locking large
momenta.

We note that it is the same free energy $F_{\rm EK}(p)$ that
causes the spontaneous breakdown of the center symmetry in the EK
model.  In that case the momenta are fully dynamical, and $F_{\rm EK}(p)$
causes them to be equal, as discussed in Sec.~\ref{EK-review}. 
This collapse is prevented by quenching, but quenching, which acts
independently in each direction, does not prevent correlations between
momenta in different directions, such as that induced by ``locking''.

We have numerically checked the argument that minimizing
$F_{\rm EK}(p)$ leads to locking in the following way. 
We considered the clock momenta,
and evaluated $F_{\rm EK}$ of \Eq{F1} for many random permutations of the momenta.
What we find is that the vast majority of permutations have a free energy
larger by $O(N^2)$ than that for the completely locked case. 
(An example of this result is given below in Fig.~\ref{F_vs_M}.)
Thus as $b\to\infty$ at fixed $N$, the completely locked vacua dominate. 
As noted in Sec.~\ref{QEK_rho_choices}, the $(Z_N)^d$ transformations and 
reflections form a  subset of the permutations, 
and for these the free-energy is invariant. Thus 
there are $(2N)^{d-1}$ degenerate vacua of the locked type,
whereas for general (non-clock) momenta we expect only a single vacuum.

In preparation for the numerical study, we now discuss quantities
that can be used to discern the predicted locking of momenta.
As we will explain, for the clock momenta these are appropriately called
order parameters, although for general $\rho(p)$ they are not.
The simplest choices are the expectation values of the $d(d-1)$ open
loops
\begin{equation}
M_{\mu,\nu} \equiv \tr(U_\mu U_\nu)/N \quad{\rm and}\quad M_{\mu,-\nu}
\equiv \tr(U_\mu U_\nu^\dagger)/N \quad (\mu>\nu)\,,
\label{eq:orderparameters}
\end{equation}
which are sensitive to correlations between gauge fields in different
directions. The utility of these quantities is particularly clear for
the clock momenta, for which one of the permutations leads to
the $\Lambda_\mu$ being equal in all directions.
Then, if $V_\mu\to 1$,
half of the $|M_{\mu,\nu}|$ equal unity ($M_{\mu,-\nu}=1$), while the other half vanish
($M_{\mu,\nu}=0$). 
The same absolute values of the $M_{\mu,\nu}$
hold for the other locked vacua obtained by
acting with $(Z_N)^d$ transformations, while $M_{\mu,-\nu}$ 
and $M_{\mu,\nu}$ switch roles under reflections.
This suggests using the combined quantity
\begin{equation}
M = \sum_{\mu<\nu} \left( |M_{\mu,\nu}| + |M_{\mu,-\nu}| \right),
\end{equation} 
as a signal for locking. 
We use both $M$ and the individual $M_{\mu\nu}$ in our numerical study.

To illustrate the utility of $M$,
we present in Fig.~\ref{F_vs_M} a scatter plot
of the normalized free energy versus $M$ for 
a large set of randomly chosen
permutations of the clock momenta and with $V_\mu=\bm{1}$.
We include the locked vacua by hand, since they are not among
those chosen randomly.
The figure indicates that the locked configurations have
free energies that are at least of $O(N^2)$ smaller than those of 
the ``unlocked'' configurations,
and have significantly larger values of $M$.
Taking the results at face value, one might be concerned that the
number of unlocked vacua might overcome their higher free-energy.
However their entropy factor is $\ln N! \sim N \ln N$, which is thus
smaller than the free-energy difference of $O(N^2)$ or greater.
Nevertheless, this plot does lead one to expect that,
for finite $N$, a range of ``partially unlocked'' states 
(present in the figure only for $N=10$) will be populated.

\begin{figure}[htb]
\centerline{
\includegraphics[width=15cm]{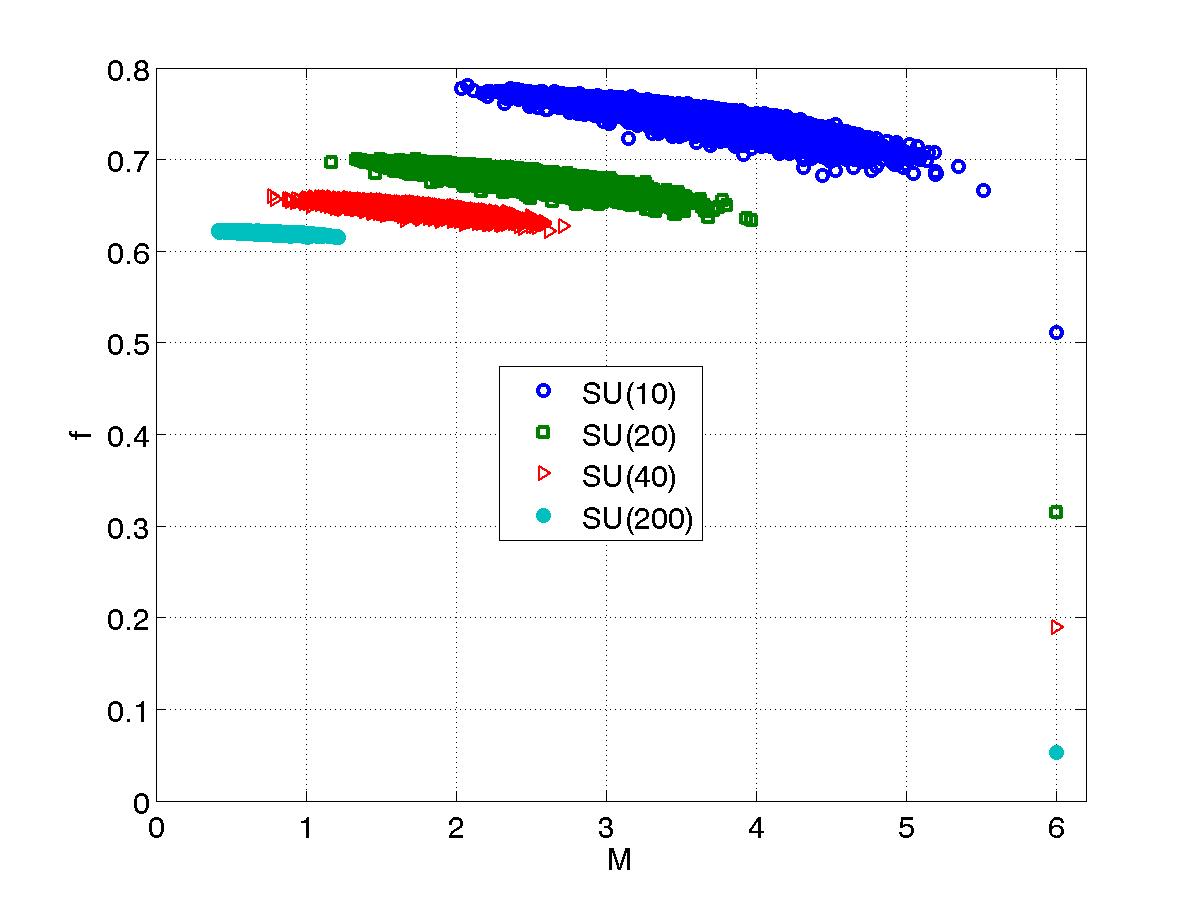}
}
\caption{The dependence of $f=F_{\rm EK}(p)/((d-2)N(N-1)/2)$ 
on the combined order parameter $M$ for random permutations of the
clock momenta. Results are for $d=4$ and $N=10,20,40,200$. 
The fully locked points with $M=6$ are included by hand. There are also 
partially unlocked states which, for each $N$, interpolate between
the mass of unlocked states and the locked ones. These do not appear
in the random sampling except for $N=10$.}
\label{F_vs_M}
\end{figure}

In Section~\ref{numerics} we use these order parameters and other numerical
evidence to argue that locking does occur also nonperturbatively.
For the remainder of this section we discuss in more detail how
locking leads to a failure of reduction.
In particular, we explain why the previous
arguments, summarized in Section~\ref{QEK_equiv}, do not hold.

\subsection{Implications of momentum locking for large-$N$ reduction}
\label{implication_of_locking}

In this section we first focus on the case of the clock momenta, and then return to other
choices.  As noted above, the quenched average in
this case is just an average over permutations of the momenta.  But we
now understand that the non-perturbative QEK model automatically includes
this sum over permutations---it is self-averaging.  Thus, in
principle, the additional quenched average is unnecessary.  We also
know, however, that the permutations are included with different
relative weights---this is manifest in the weak coupling free-energy
landscape of Fig.~\ref{F_vs_M}, and there is no reason to expect
equality for other couplings. Regardless of the details, the mere
fact that the weights are different implies that the integrations over
momentum space that are induced by the sums over color indices are
not uniform.  This is sufficient to invalidate reduction 
---the momentum integrations in the reduced and
infinite-volume cases are different.  The case of complete locking
provides an extreme example: the momentum of each gluon then has the
same component, $\Delta p_\mu^{ab}$, in each direction, and the
integration over the $d$-dimensional Brillouin zone collapses to an
integration along the 1-dimensional body diagonal.

The argument for reduction based on the loop equations also fails,
because one or other of the key steps,
eqs.~(\ref{eq:loop1}) and (\ref{eq:loop2}), does not hold.
To see how this works, we write
out these relations for the case that $W_{\rm open}= M_{\mu,\nu}$ and
$W'_{{\rm open}}= M_{\mu,\nu}^*$:
\begin{eqnarray}
\int dp\,\,  \left\langle M_{\mu,\nu} \,\,
M_{\mu,\nu}^*\right\rangle_p\, 
&=&
\int dp\,\,\left\langle M_{\mu,\nu} \right\rangle_p\,\,
\left\langle M_{\mu,\nu}^*\right\rangle_p + O(1/N^2) 
\label{Mloop1} \\
\int dp\,\,\left\langle M_{\mu,\nu} \right\rangle_p\,\,
\left\langle M_{\mu,\nu}^*\right\rangle_p
&=&
\int dp\,\,  \left\langle M_{\mu,\nu} \right\rangle_p
\int dp' \left\langle M^\star_{\mu,\nu}\right\rangle_{p'}
+ O(1/N).
\label{Mloop2}
\end{eqnarray}
We now argue that, if locking occurs, then, for some
$\mu$ and $\nu$, the following two statements are correct :
\begin{enumerate}[(I)]
\item   The r.h.s. of
\Eq{Mloop2} is of $O(1/N)$.
\item The l.h.s. of \Eq{Mloop1} is of $O(1)$.
\end{enumerate}
Thus one or both of the relations must be wrong.
The numerical evidence of Section~\ref{numerics} suggests that it is
the second relation, \Eq{Mloop2}, which fails.

It is easy to see that statement (I) is correct 
regardless of the choice of $\rho(p)$. This is due to
the center symmetry (\ref{eq:momentaZN}), under which
\begin{equation}
\left\langle M_{\mu,\nu}\right\rangle_{p} 
\longrightarrow
\left\langle M_{\mu,\nu}\right\rangle_{p + {2\pi n}/{N}} 
=
 \left\langle M_{\mu,\nu}\right\rangle_{ p} \, e^{2\pi i  (n_\mu+n_\nu)/N}, 
\label{Mphase} 
\end{equation}
Since the measure $dp$ is unchanged when $p\to p +2\pi n/N$, the phase
factors will cause $\int dp \left\langle M_{\mu,\nu}\right\rangle_{p}$
to vanish.  For the clock momenta the situation can be slightly
different. There, self-averaging may take place, and this means that
the momenta that contribute to $\left\langle
M_{\mu,\nu}\right\rangle_{p}$ are all those related to $p$ by
permutations. These include also the momenta $p^a_\mu+2\pi n_\mu/N$
with $n_\mu$ integer, and so $\left\langle
M_{\mu,\nu}\right\rangle_{p}\sim \sum_{n_\mu, n_\nu} e^{2\pi i
(n_\mu+n_\nu)}=0$. Consequently we see that self-averaging in the
clock momenta case makes statement (I) correct even without the
integrations.

To see why statement (II) is correct note that the locking means that
some of the $|M_{\mu,\nu}|$ will have $O(1)$ values. In contrast to the integrands of \Eq{Mloop2}, the integrand here, $M_{\mu,\nu} M_{\mu,\nu}^*=|M_{\mu,\nu}|^2$, is invariant under the
center symmetry, and thus maintains its $O(1)$ value even after 
integration.

Which of the two Equations~(\ref{Mloop1}) and~(\ref{Mloop2}) fails? 
This depends on the nature of the dynamics.
If the self-averaging occurs, then the second step, which for clock
momenta is simply
\begin{equation}
\left\langle M_{\mu,\nu} \right\rangle_p\,\,
\left\langle M_{\mu,\nu}^*\right\rangle_p 
=
\left\langle M_{\mu,\nu} \right\rangle_p
\left\langle M^\star_{\mu,\nu}\right\rangle_{p'}\,,
\end{equation}
is trivially valid, and it is the first step which fails.
This breakdown of large-$N$ factorization is then an example
 of the breakdown of cluster decomposition due the presence 
of multiple vacua ---all those related by the center and reflection symmetry.

The other possibility is spontaneous symmetry breaking (SSB)
of the center symmetry, in which the
system gets stuck in the vicinity of one of the locked vacua.
We recall that $Z(p)$ itself is $(Z_N)^d$ symmetric with clock
momenta, so there is a symmetry to break.
Furthermore, despite the fact that the QEK model has zero volume,
SSB is possible when $N\to\infty$
because there are then an infinite number of degrees of freedom.
The $\left\langle M_{\mu,\nu}\right\rangle_p$ (or, indeed, the
quenched expectation values of any open loops) are order 
parameters---non-vanishing values indicate SSB.
If SSB takes place then, by definition, self-averaging no longer occurs, 
and vacuum expectation values of open loops
 vanish only if we explicitly average over the input permutations.
If the input permutation is changed by a center transformation, then,
since the dynamics is $(Z_N)^d$ invariant,
the vacuum that is selected will also be changed by the same
transformation.
In this possibility, factorization, \Eq{Mloop1}, holds, 
because fluctuations about the single vacuum are suppressed as $N\to\infty$.
It is the second step, \Eq{Mloop2}, that fails. On the l.h.s. the
same vacua are selected in the two quenched expectation values,
because the same input momenta are used,  while on the r.h.s. different
vacua are, in general, selected. Thus the l.h.s. will be of $O(1)$ for
all input $p$, while the first term on the r.h.s. will average to zero.
Thus what we call quenched factorization fails.

We discuss which of the two possibilities---self-averaging or
SSB---is expected to occur in the
next subsection. Regardless of which occurs, however, the key
point is that the combination of the relations (\ref{Mloop1}-\ref{Mloop2})
fails, either invalidating cluster decomposition or breaking the center symmetry, and thus large-$N$ reduction fails. Furthermore,
one can numerically test for this by calculating the l.h.s. of
(\ref{Mloop1}) and determining whether it falls as $1/N$ (as required
for reduction) or tends to a constant as $N\to\infty$ (reduction fails).

\bigskip

We now consider the uniform weight function, $\rho(p)=1$.
In this case $Z(p)$ is not center-symmetric and the $M_{\mu,\nu}$ are
not order parameters. Nevertheless, if locking occurs, we
expect something similar to SSB to take place. For a random input
choice of $p$, we expect the system to sample the space of
permutations, until it finds that with the smallest free energy.
Note that none of the permutations will be related by center
or reflection symmetries, so all are expected to have different
free energies. In this picture, the system ends up fluctuating
in the vicinity of a particular permutation. If the weak-coupling
free energy is any guide, the chosen permutation will be such
that if, for a given pair of indices $(a,b)$, the difference
$\Delta p_\mu^{ab}$ is small for one value of $\mu$, then it will
also be small for all other values of $\mu$.
In other words the chosen momenta will be partially locked.\footnote{%
The complete locking possible for clock momenta is not possible here
because the components of $\Delta p$'s in different directions are
different.}
Thus, even though the input momenta are uniformly distributed
in the BZ, those chosen dynamically are not,
and planar perturbation theory is not correctly reproduced.

The expected partial locking implies that, for most input
$p$, some of the $M_{\mu,\nu}$
will fluctuate around complex values with magnitudes of $O(1)$. 
If so, this invalidates the quenched factorization of \Eq{Mloop2},
because the l.h.s. averages to an $O(1)$ value, while the
$p$-integrals on the r.h.s. implement the center symmetry
and cause the averages to vanish. This picture is confirmed
by our numerical findings in Section~\ref{numerics}.

For $\rho_{\rm BZ}$ the situation is
similar to that for $\rho_{\rm clock}$.
There are multiple locked vacua
related by the center and reflection symmetries, and locking
invalidates reduction. The difference
is that it is only the $(Z_K)^d$ subgroup of the full center symmetry
which is realized, where $K=N^{1/d}$.
To make clear how the presence of permutations in the dynamics unravels
the carefully chosen coverage of the BZ, we can refer to the 
simple example in Fig.~\ref{BarsN16d2}. Permuting the momentum components in the ``1'' direction
as, for example, $p_1^a\leftrightarrow p_1^b$ for $(ab)=(25)$, moves the
two momenta in the ``boxes'' labeled 2 and 5 in the Figure into
those labeled by 1 and 6, where the momenta are locked. 
Similarly, all other off-diagonal pairs
can be moved by permutations onto the diagonal. Thus if the free-energy
favors locking, as the weak-coupling argument implies, then the momenta
chosen by the simulation will lie on the one-dimensional diagonal of the BZ.\footnote{%
Here we note again that, when any of the $p_\mu^a$ are equal, as 
they are for $\rho_{BZ}(p)$, then there are flat directions which are
not Gaussian, and the form $F_{EK}(p)$ of \Eq{eq:F2} is invalid. As
mentioned above, we do not study further the effect of these flat
directions, but rather investigate the QEK model with Monte-Carlo
simulations (see next section).}

Finally, we briefly discuss the choice $\rho(p)=\rho_{\rm VdM}(p)$ of \Eq{eq:Weylweight}.
This is in some sense intermediate between the clock
and uniform choices. On the one hand, any value of $p$ is possible
with $\rho_{\rm VdM}(p)$, while, on the other,
the large-$N$ limit of $\rho_{\rm VdM}(p)$ is 
$\rho_{\rm clock}(p)$. Thus we expect locking or partial-locking for
$\rho_{\rm VdM}$, and this is indeed what we find numerically.

\subsection{Expected size of fluctuations}

In this subsection we address the question of whether, for the
clock momenta, we expect the theory to exhibit SSB or not.
We are interested in this question for fixed $b$ and $N\to\infty$.
The weak-coupling result of \Eq{F1} provides a guide
to the free-energy landscape, and suggests that the dominant states correspond
to fluctuations about the locked vacua.
As in any statistical mechanical system, the issue is whether the
fluctuations are large enough to cause the theory to move from one
locked vacuum to others related by symmetry transformations.
In infinite volume, we know
from the Mermin-Wagner-Coleman theorem \cite{MWC} that for $d>2$
the fluctuations are not IR divergent and SSB is possible,
while for $d\le2$ it is not.
The question is how this result translates to the QEK model where the
spatial volume is embedded in the color space.

To get a rough idea of what happens, imagine that we are in a completely
locked vacuum.
A measure of the fluctuations in the (normalized) traces of
open Wilson loops (such as the $M_{\mu,\nu}$) 
is given by the ``tadpole'' graph
\begin{equation}
T 
\equiv 
\frac{g^2}{N} \sum_{a\ne b} \langle A_\mu^{ab} A_\mu^{ba}\rangle_{p}
= 
\frac{g^2}{N} \sum_{a\ne b} \frac4{\sum_\nu \sin^2(\Delta p_\nu^{ab}/2)} \,,
\label{eq:tadpole1}
\end{equation}
where $\mu$ is fixed, and we have used \Eq{gluon_prop}.
The $g^2$ comes from expanding the $U_\mu$, and the $1/N$ from the normalized
trace. Note that since we are doing perturbation theory we can really fix
the momenta, and we are taking $p$ to be locked.
This means that $|\Delta p_\nu^{ab}|=|P^a-P^b|$ is independent of $\nu$,
and the tadpole can be rewritten as
\begin{equation}
T_{\rm locked} = \frac{g^2}{N} \sum_{a\ne b} 
\frac1{\sin^2([P^a-P^b]/2)} 
= 1/b \int_{c/N}^{\pi} \frac{dq}{\pi} \frac1{\sin^2(q/2)} + O(1/N)\,.
\end{equation}
As $N\to\infty$, the sum has gone over to an integral,
but the integral is over a {\em one-dimensional} momentum space,
and is thus IR divergent. The cut-off $c/N$
(with $c$ a constant that could be determined by a more 
complete analysis) arises from fact that the original sum,
\Eq{eq:tadpole1}, has a minimum $\Delta p$ of $O(1/N)$.
The IR divergence implies that
\begin{equation}
T_{\rm locked} \propto  N/b \,[1 + O(1/N) ]
\label{eq:tadpole_div}
\end{equation}
so that the fluctuations about the locked vacua diverge as $N\to\infty$
for fixed $b$. 

These divergences can be anticipated from the result for the maximum
energy barrier $\Delta S_{\rm QEK}$ (see
Eqs.~(\ref{eq:transp_V})-(\ref{eq:Parsonsbarrier})) that exists
between two configurations related to each other by the permutation
$p^a_\mu \leftrightarrow p_\mu^b$. Denoting $\Delta p_\mu^{ab}\equiv
p^a_\mu - p_\mu^b$, we see that if $\Delta p_\mu^{ab}=O(1/N)$ and
$\Delta p^{ab}_\nu \sim O(1)$, then $|\Delta S_{\rm QEK}| \sim b/N$,
and fluctuations in the direction parameterized by the $SU(N)$ matrix
(\ref{eq:transp_V}) overcome the barrier when $N$ is large enough 
that $b/N < O(1)$. 
For locked vacua with $\Delta p_\mu=O(1/N)$ in all
directions, the barrier is even lower, $|\Delta S_{\rm QEK}|\sim b/N^3$.
 The Gaussian terms in the action for these
``flat'' directions is small relative to higher-order terms, and
ignoring the latter in the tadpole calculation leads to the apparent
IR problem. 

It is, in fact, the more severe $b/N^3$ divergence
which leads to $T_{\rm locked}\propto N$. One can see this
by noting that for a random permutation of the clock momenta
$T \sim \int d^4 q/q^2 $ in the IR, and this is convergent.
This is despite the fact that there are the flat directions
with $|\Delta S_{\rm QEK}| \sim b/N$. 

The upshot of this discussion is that we cannot
quantitatively trust the weak-coupling calculation of $T$
for the locked vacua if $N\to \infty$ at fixed $b$. 
How does this affect the free-energy $F_{EK}(p)$ which
we discussed above for the locked vacua?
It follows from \Eq{F1} that 
\begin{equation}
F_{\rm EK}(p_{\rm locked}) \sim N^2 \int_{c/N}^{O(1)}  dq\, q \log(q) 
\sim O(N^2) + O(N\log N) \,,
\end{equation}
and so the leading order term is IR safe, while the
subleading term cannot be trusted in a Gaussian analysis.
Since our previous discussion was based on the leading
$O(N^2)$ term, it remains valid.

Returning to the issue of SSB, we need to know whether
the large-$N$ divergence of $T_{\rm locked}$ implies that
the system will fluctuate into nearby locked vacua (which
have momenta differing by center-transformations or reflections).
We know that there will be large fluctuations in the directions
given by transpositions between close momenta, for these are
the source of the IR divergence.
Thus to address the question we proceed as follows.
It is possible to move from one locked vacuum to another
by stringing together a sequence of transpositions involving 
nearby $p$'s (i.e. with $\Delta p$ always of $O(1/N)$).
As we proceed along such a string, the momenta become partially unlocked,
and the energy barriers to transpositions increase from of $O(1/N^3)$ to
of $O(1/N)$. Nevertheless, they
all still vanish as $N\to\infty$, so there is a vanishing energy barrier
between locked vacua. What matters, however, is whether
there is a free-energy barrier. We can investigate this using the
weak-coupling result, evaluating \Eq{F1} numerically for each
momenta along the path.

\begin{figure}[htb]
\centerline{
\includegraphics[width=10cm,clip=true]{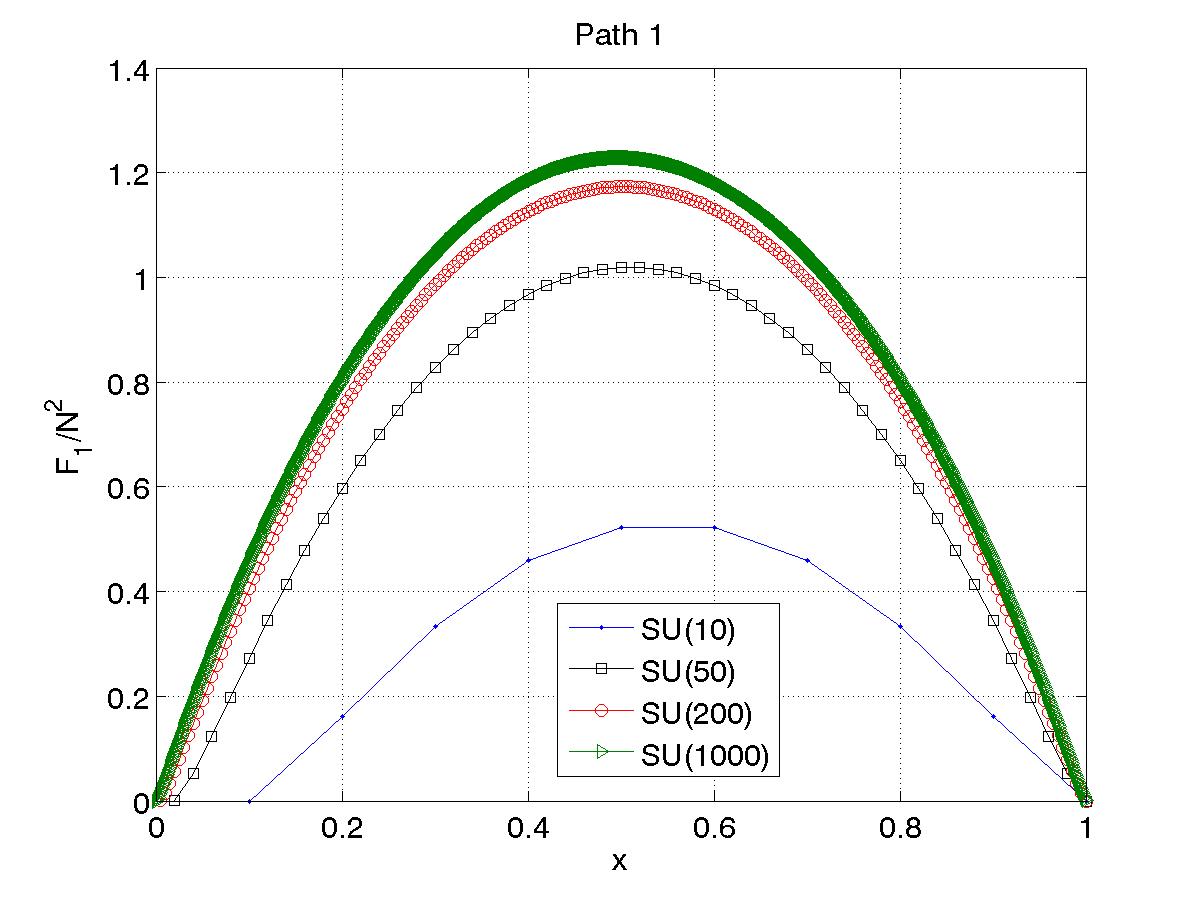}
\includegraphics[width=10cm,clip=true]{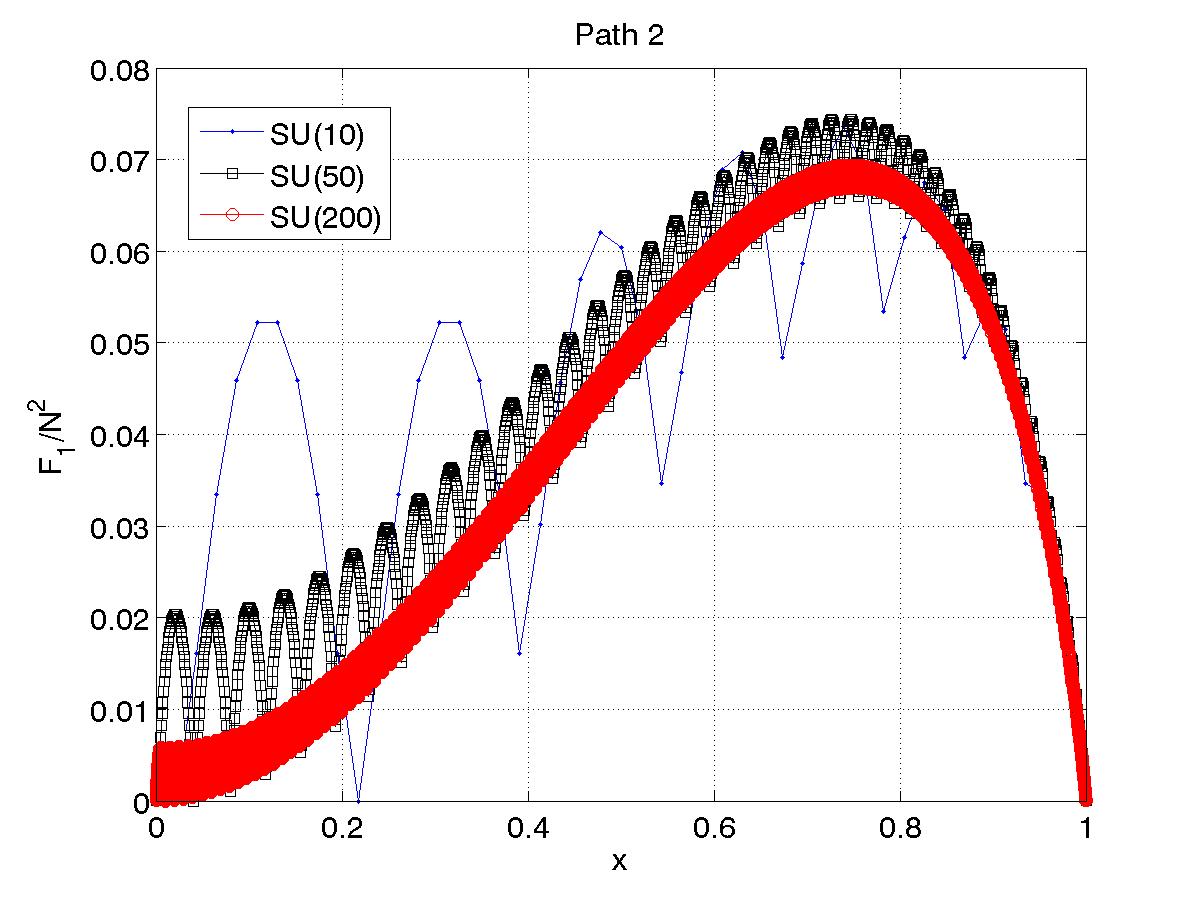}
}
\caption{The free energy $F_{\rm EK}(p)/(d-2)$, divided by $N$ (left panel)
or $N^2$ (right panel),
along the two paths between locked vacua described in the text.
Results are for $d=4$ and $N=10$, $50$, $200$ and $1000$.
The horizontal axis gives the fraction of the total transpositions
required, with the starting point being plotted at position $1/N$ (left panel)
and $2/[N(N-1)+1]$ (right panel).
The left panel is for path 1 (vacua related by center-symmetry), the
right panel for path 2 (vacua related by a reflection).}
\label{F_paths}
\end{figure}

We have considered two types of path.
Both begin from a locked state in which $U_\mu=\Lambda_1$ for all $\mu$,
where $(\Lambda_1)_{aa}=\exp(iP^a)$, with the $P^a$ given in
\Eq{eq:clockmomenta}.
Thus, for example, $M_{\mu,\nu}=0$ and $M_{\mu,-\nu}=1$. Path 1 arrives at
a state with $U_1=U_3=U_4$ unchanged and $U_2=U_1 e^{-2\pi i/N}$,
so that $M_{12}=0$, $M_{1,-2}=e^{+2\pi i/N}$, etc.. 
The path is made of a series of $N-1$ transpositions between adjacent indices, 
for each of which $|\Delta p|$ takes its minimal value of $2\pi/N$.
The paths we use are exemplified by the following sequence
for $N=6$ (which shows the ordering of the momenta $P^a$, $a\in [1,6]$,
along the diagonal of $U_2$)
\begin{equation}
123456 \to 213465 \to 312564 \to 412356 \to 512346 \to 612345 \,.
\label{path1}
\end{equation}
In contrast, path 2 takes us to a vacuum with
$U_1=U_3=U_4=\Lambda_1$ and $U_2=\Lambda_1^\dag$, for which 
$M_{1,2}=1$ and $M_{1,-2}=0$.
This can be achieved with transpositions alone along a more
complicated path of length $N(N-1)/2$, 
that for $N=6$ would be the string of transpositions in \Eq{path1}, followed by
\begin{eqnarray}
&612345&\to 6213465 \to 631245 \to 641235 \to 651234
\to 652134 \to 653124 \to 654123  \to \nonumber \\ 
&654213& \to 654312 \to 654321.
\end{eqnarray}
We show the results in Fig.~\ref{F_paths}. 
We find that, for path 1, the free energy
barrier $\Delta F$ scales asymptotically
with $N$, while for path 2 it scales with $N^2$.
Since we know from above that the $O(N^2)$ part of the free energy is IR safe,
we conclude that fluctuations along path 2 are certainly suppressed.
For path 1 the situation is more subtle, as we now discuss.

The issue for path 1 is whether the leading contribution 
to $\Delta F$, which we see to be of $O(N)$, is IR safe,
given that $F$ itself is untrustworthy at this order.
The numerical results themselves suggest that the $N\log N$ terms cancel
in $\Delta F$, but it would require a more detailed analytic
analysis to demonstrate that this cancellation of untrustworthy
terms is itself trustworthy. Thus the most conservative conclusion
is that we do not know whether the barrier path 1 grows with $N$
and suppresses fluctuations.
Other uncertainties in this analysis are that
we have not investigated all paths, nor accounted
for a possible entropy factor involving the number of paths,
and finally that it is based on the leading term in the weak-coupling
analysis. Thus to learn about the extent of locking, and
the possibility of SSB, we must study the QEK model non-perturbatively,
and to this we now turn.

\section{Non-perturbative Lattice Study}
\label{numerics}

In this section we present our numerical results for
the QEK model. 
In Sec.~\ref{setup} we briefly describe the methodology,
focusing on an explanation of the two strategies we adopt
to perform the quenched average: self-averaging and
explicit averaging. 
In Sec.~\ref{Map} we map the phase structure 
as a function of the bare coupling, $b$,
using measurements of the plaquette and the
order parameters $M_{\mu,\nu}$. These results lead us to investigate
various features of the model in more detail. In
Sec.~\ref{more_precise_plaquette} we describe results
from high-precision
measurements of the plaquette. This allows us to study the
dependence of the results on the calculational strategy and on the
choice of the weight function $\rho(p)$ (defined in Section~\ref{QEK_equiv}).
We present similar measurements for $M_{\mu,\nu}$ in
Sec.~\ref{more_precise_M} and use them to understand the structure
of the vacua of the QEK model. Finally, in Sec.~\ref{bt_w_WL}, we
analyze a ``strong-to-weak'' transition that occurs in the model,
using an adaptation of the
Wang-Landau algorithm~\cite{WL} to perform a precise measurement
of the coupling $b_t$ at which it occurs.

\subsection{Methodology}
\label{setup}

The QEK model has been defined above in 
eqs.~(\ref{eq:O(U)theta})-(\ref{eq:QEKexpectationvalue}).
The ingredients for a simulation are a weight function for
the momenta, $\rho(p)$, and the coupling $b$ in the quenched action,
eq.~(\ref{eq:SQEK}). Specifically
one is instructed to draw momenta weighted
by $\rho(p)$, construct the  diagonal eigenvalue matrices
$\Lambda_\mu$ using eq.~(\ref{eq:momentadef}),
and then do a Monte-Carlo average over the $SU(N)$ matrices
$V_\mu$ for fixed $\Lambda_\mu$. Observables involving gauge
links, such as the plaquette, can then be reconstructed using the
definition $U_\mu=V_\mu^\dagger \Lambda_\mu V_\mu$.

As noted above, we consider four choices of weight function:
uniform ($\rho_{\rm uniform}(p)=1$),
clock [defined in eq.~(\ref{eq:rho_clock_def})],
Vandermonde [defined in eq.~(\ref{eq:Weylweight})],
and BZ [defined above eq.~(\ref{Bars_p})].
It is straightforward to draw momenta from the first three
of these distributions, while there is only a single choice for
the BZ distribution.

The Monte-Carlo integration over the $V_\mu$ is non-standard
because the action (\ref{eq:SQEK}) is quartic in each of these matrices,
so that a simple heat-bath algorithm cannot be used. 
Instead, we use the following three approaches.
\begin{enumerate}
\item 
A straightforward (and slow) Metropolis algorithm using the original action,
updating all of the $SU(2)$ subgroups of each $V_\mu$ in turn;
\item
A faster Metropolis algorithm, using
a Gaussian auxiliary field to reduce the
action to quadratic order in the $V_\mu$~\cite{FabriciusHahn}.
We again update $SU(2)$ subgroups of $V_\mu$ in turn.
\item 
A combination of a Cabibbo-Marinari heat-bath 
(again applied in turn to $SU(2)$ subgroups) 
and various type of over-relaxations
(both $SU(2)$ and $SU(N)$, the latter using
the method of Ref.~\cite{SUNOR}). These are applied after using two
Gaussian auxiliary fields to make the action linear in the $V_\mu$. 
We use a ratio of one heat-bath update for every four over-relaxations.
The details of this algorithm are described in Ref.~\cite{WL_paper}. 
\end{enumerate}
We find that the second algorithm typically decorrelates our measured
quantities most rapidly, and we use this for most of our runs.
All the measurements in this paper were separated by 5 full updates of all
four $V_\mu$'s. 

We perform the evaluation of the quenched average,
\Eq{eq:QEKexpectationvalue}, using one of the
following two strategies. The only exception is
for $\rho_{\rm BZ}$, for which no average is necessary.

\vskip 0.5cm
\noindent
\underline{Strategy A : \quad Explicit quenched averaging}

\vskip 0.5cm
\noindent
Here we simply follow the quenching recipe laid out above:
generate an ensemble
of sets of momenta weighted by the distribution $\rho(p)$, calculate
$\langle{\cal O(U)}\rangle_p$ for each member of this ensemble, and then
average over the ensemble. We have analyzed the QEK with this strategy
for all choices of $\rho(p)$ listed above,
except $\rho_{BZ}$, but have mostly focused on
$\rho_{\rm clock}(p)$.

\vskip 0.5cm

\noindent
\underline{Strategy B: \quad Self-averaging}

\vskip 0.5cm

\noindent
As mentioned in Sec.~\ref{QEK_equiv}, if reduction
is valid, then one need not 
perform the momentum integral at large-$N$---a single
value $p=p_0$ is sufficient, since the sum over color
indices will sample the Brillouin zone. 
We refer to this possibility as self-averaging.
For the choice $\rho(p)=\rho_{\rm clock}(p)$, self-averaging
is, in principle, exact for all finite $N$, because, as explained 
in Sec.~\ref{QEK_rho_choices},
\begin{equation}
\langle{\cal O(U)}\rangle_{\rm QEK} 
= \langle{\cal O(U)}\rangle_{p_0} \,.
\label{eq:self_averaging_clock}
\end{equation}
Here, for each $\mu$,  $(p_0)_\mu$ can be any permutation 
of the clock momenta $P_a$ defined in \Eq{eq:clockmomenta}.
We recall that \Eq{eq:self_averaging_clock}
holds because the integral over $V_\mu$ includes all 
permutations of the elements of $\Lambda_\mu$. 
For $\rho=\rho_{\rm clock}$, 
we often use $(p^a_\mu)_0=P^a$ for all $\mu$
(which we call $p_0=p_{\rm locked}$).

The self-averaging strategy 
is not guaranteed to work in practice, because it may
be that the simulation fails to fully explore all possible
permutations due to algorithmic shortcomings~\cite{Parsons,KNN}. 
To check whether this happens it is
useful to measure quantities which change as one moves
from one permutation to another, and we use the
order parameters $M_{\mu,\nu}$ [\Eq{eq:orderparameters}]
for this purpose.
 
It is important, however, to distinguish such an algorithmic
failure from a genuine breakdown of reduction.
In the latter case, most of the permutations will
have higher free energy, and will be visited with a probability
which vanishes as $N\to\infty$. 
Furthermore, if spontaneous symmetry breaking occurs (as is possible
with $\rho_{\rm clock}$ or $\rho_{\rm BZ}$) then, as $N\to\infty$,
the system  fluctuates around a single vacuum due
to the infinite barrier between vacua connected by particular permutations.

\subsection{Mapping $b$ dependence}
\label{Map}

We begin by determining the dependence of the average plaquette,
\begin{equation}
u_p \equiv \frac{1}{Nd(d-1)/2} \, \Re \sum_{\mu>\nu} \< \Tr U_\mu U_\nu
U^\dag _\mu U^\dag_\nu \>\,,
\end{equation}
and of the $M_{\mu,\nu}$,
as a function of $b$ in the range 
$0.1 \stackrel{<}{_\sim} b \stackrel{<}{_\sim}1$. 
We use the self-averaging strategy, and 
have used all choices of $\rho(p)$.
Our focus in this section is on qualitative features,
and at this level we do not find much dependence
on the choice of $\rho(p)$. For brevity, therefore, we only
present results for $\rho(p)=\rho_{\rm clock}$.\footnote{%
We do see, however, that
the $1/N$ corrections for $\rho_{\rm uniform}(p)$ are very
large for $b\stackrel{<}{_\sim}0.3$ and smear a strong-to-weak
transition that occurs at about $b\simeq 0.32$. This observation was
also reported in Ref.~\cite{Okawa2}.} 

We use ``hysteresis runs''
starting either from a ``cold'' field configuration with
\begin{equation}
V_\mu = \bm{1}\,, 
\label{cold_start}
\end{equation}
at a high value of $b=b_{\rm cold}\simeq 0.7-1.0$, 
or from a ``hot'' field configuration with
\begin{equation}
V_\mu = {\rm random \,\,element \,\,of \,\,} SU(N)\,,
\label{hot_start}
\end{equation}
at a low value of $b=b_{\rm hot}\simeq 0.1-0.2$. 
For a cold (hot) start we gradually decrease (increase) $b$ 
until we reach the value $b=b_{\rm hot (cold)}$. 
We study gauge groups with $10\le N\le 200$, and 
list the parameters of our major runs in
Table~\ref{detailsMap}.

\begin{table}
\setlength{\tabcolsep}{4mm}
\begin{tabular}{cccc}
\hline\hline
$N$ & equilibration updates  & measurements & $\rho(p)$ and $p$\\
\hline 
$20$, $40$ & $100$ & $200$ & uniform, VdM \\ 
$20$, $40$ & $1000$ & $5000$ & clock ($p_{\rm locked}$)\\ 
$50$ & $1000$ & $5000$ & clock ($p_{\rm locked}$) \\ 
$80$ & $100$ & $200$ & $\left\{\parbox{3cm}{\vskip 0.2cm uniform \\VdM
\\clock ($p_{\rm locked}$) \vskip 0.2cm }\right.$ \\ 
$100$ & $500$ & $1000$ & clock ($p_{\rm locked}$) \\ 
$16$ & $1000$ & $5000$ & $BZ$ \\ 
$81$ & $100$ & $500$ & $BZ$ \\
\hline\hline
\end{tabular}
\caption{Details of hysteresis runs used to map the phase diagram.
For each value of $b$, we use the quoted number of equilibration
updates followed by the quoted number of measurements
(the latter being made every 5 updates).}
\label{detailsMap}
\end{table}

\begin{figure}[htb]
\includegraphics[width=12cm]{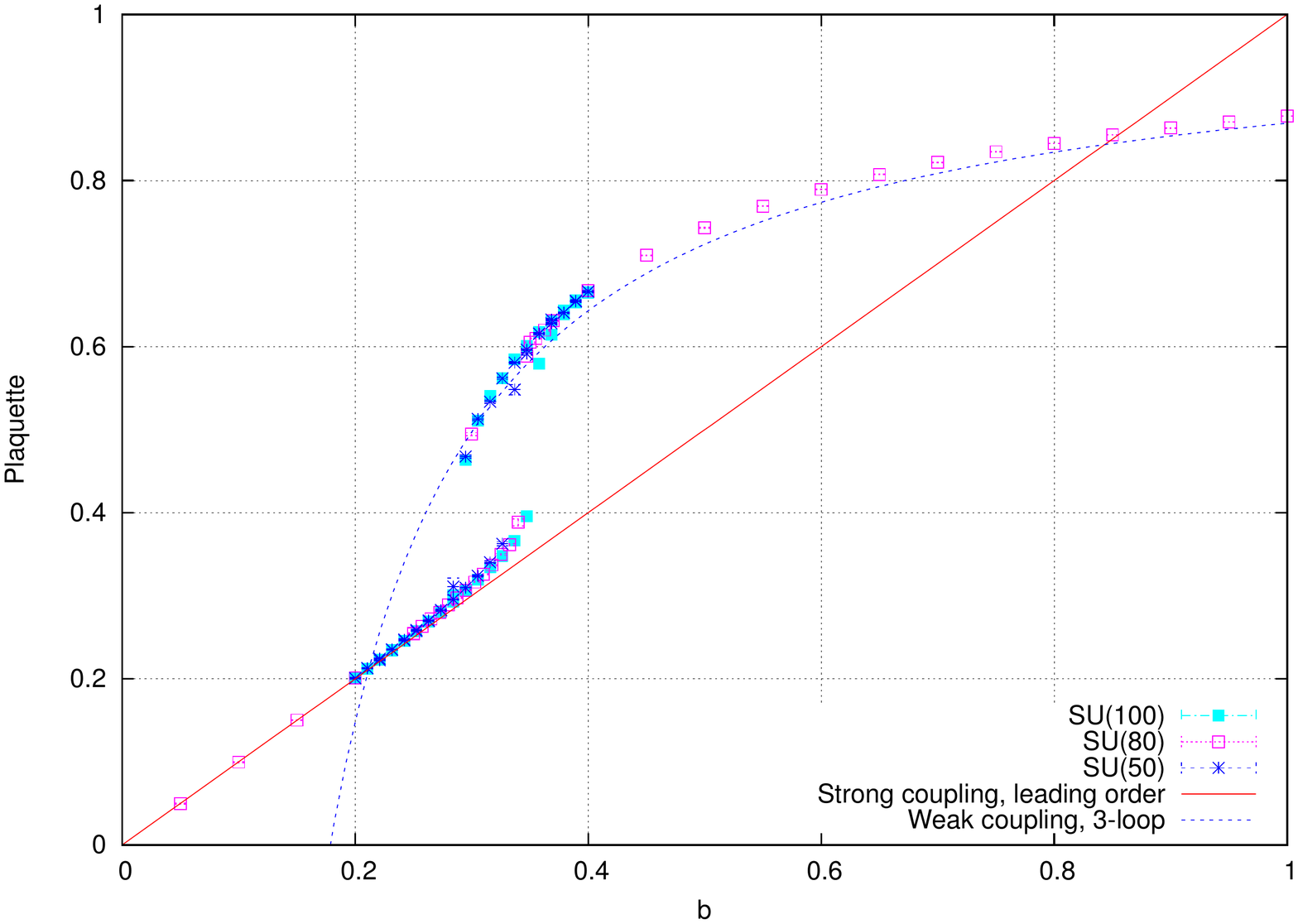}
\\
\includegraphics[width=12cm]{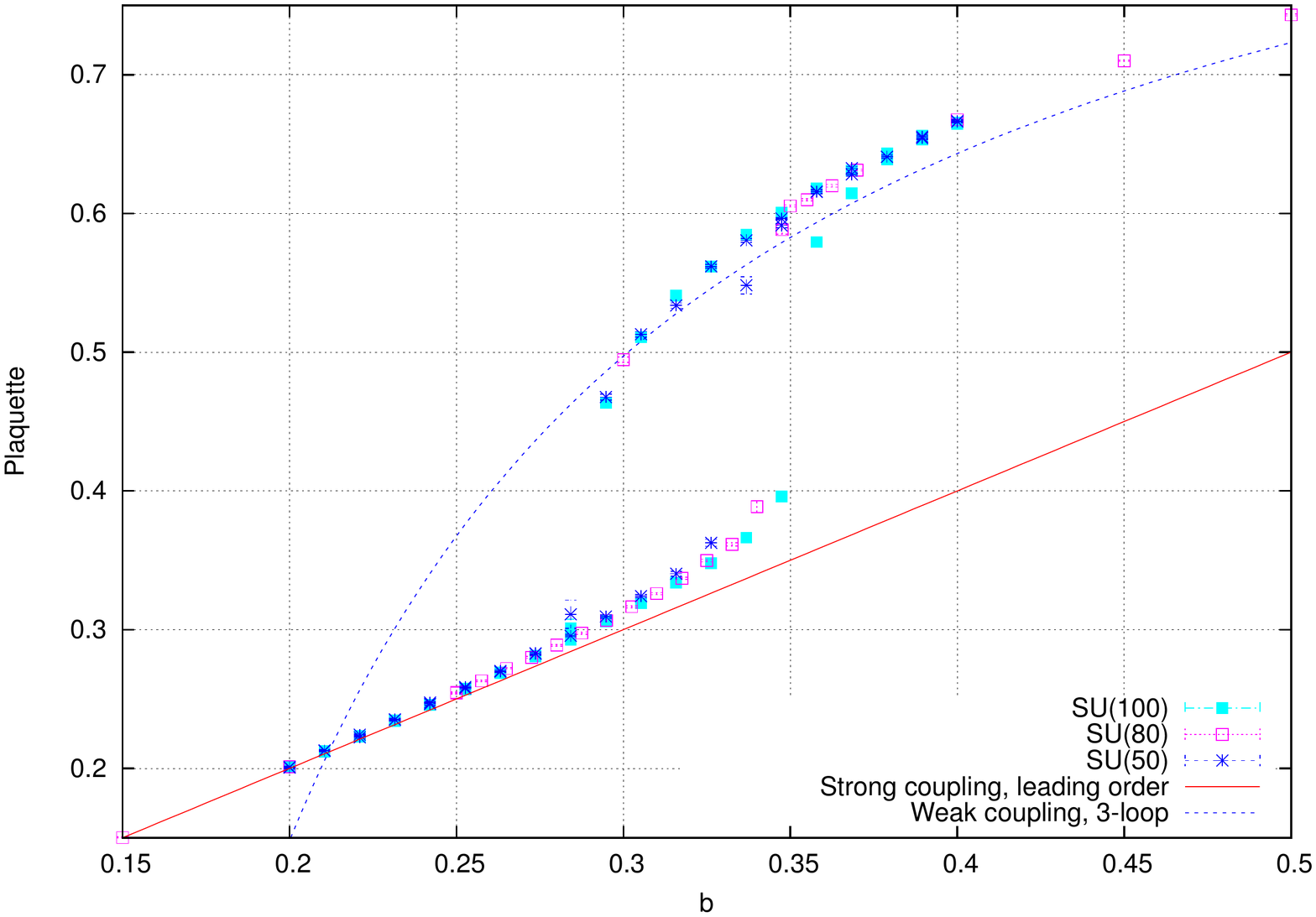}
\caption{Hysteresis plots of the plaquette variable $u_p$ versus $b$
for $SU(50)$ ([blue] crosses), 
$SU(80)$ ([magenta] open squares), 
and $SU(100)$ ([light blue] filled squares). 
Results are for $\rho_{\rm clock}$ and self-averaging.
The curves are the predictions for $SU(\infty)$ in
the strong-coupling expansion to leading order 
(solid [red] curve) and of the weak-coupling
expansion to three loop order (dashed [blue] curve)
(taken from, for example, Ref.~\protect\cite{TEK}).
The lower-panel shows a close-up of the 
strong-to-weak transition region.}
\label{hyst_map}
\end{figure}

In Fig.~\ref{hyst_map} we present results for $u_p$
from simulations with $N=50$, $80$ and $100$.
At first glance, the plots appear consistent with
the validity of reduction. The
results for $u_p$ are close to the analytic predictions and to
the numerical results from large volume, large-$N$ simulations (although we do not show the latter here). 
The increasing hysteresis with increasing $N$ is indicative
of a strongly first order phase transition somewhere in the
range $b_t\simeq 0.30-0.35$, 
which is indeed close to the coupling, $b_{\rm bulk}\simeq 0.36$,
at which the well-known ``bulk'' transition 
occurs in the large-$N$ gauge theory~\cite{Mike}.\footnote{%
There is also an estimate of $b_{\rm bulk}$ from 
the TEK model~\cite{Campostrini}, the equivalence of
which to large-$N$ pure gauge theory has been thrown
into doubt by the work 
of Refs.~\cite{TV,Ishikawa,Bietenholz}.}

\begin{figure}[htb]
\includegraphics[width=13cm]{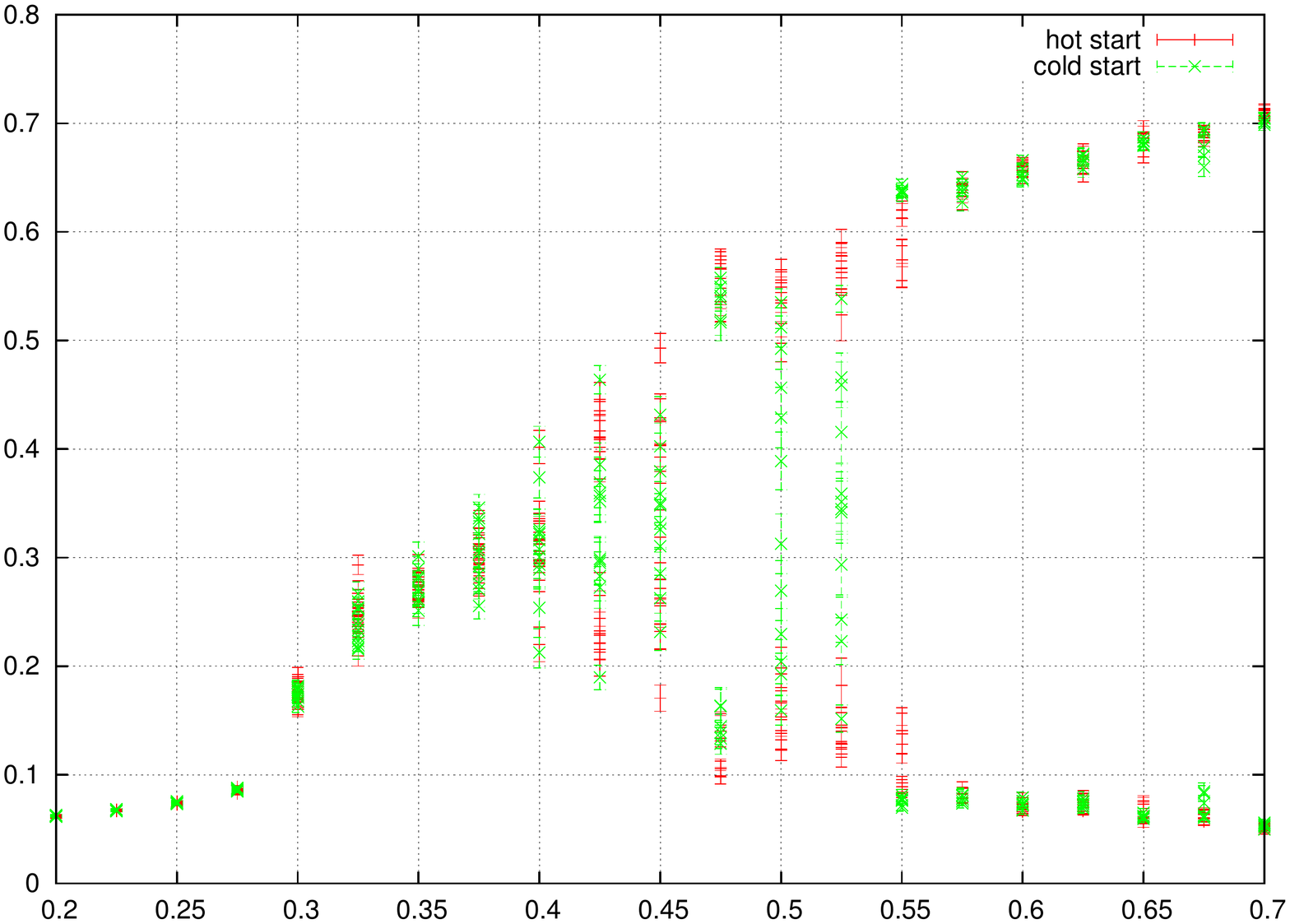}
\\
\includegraphics[width=13cm]{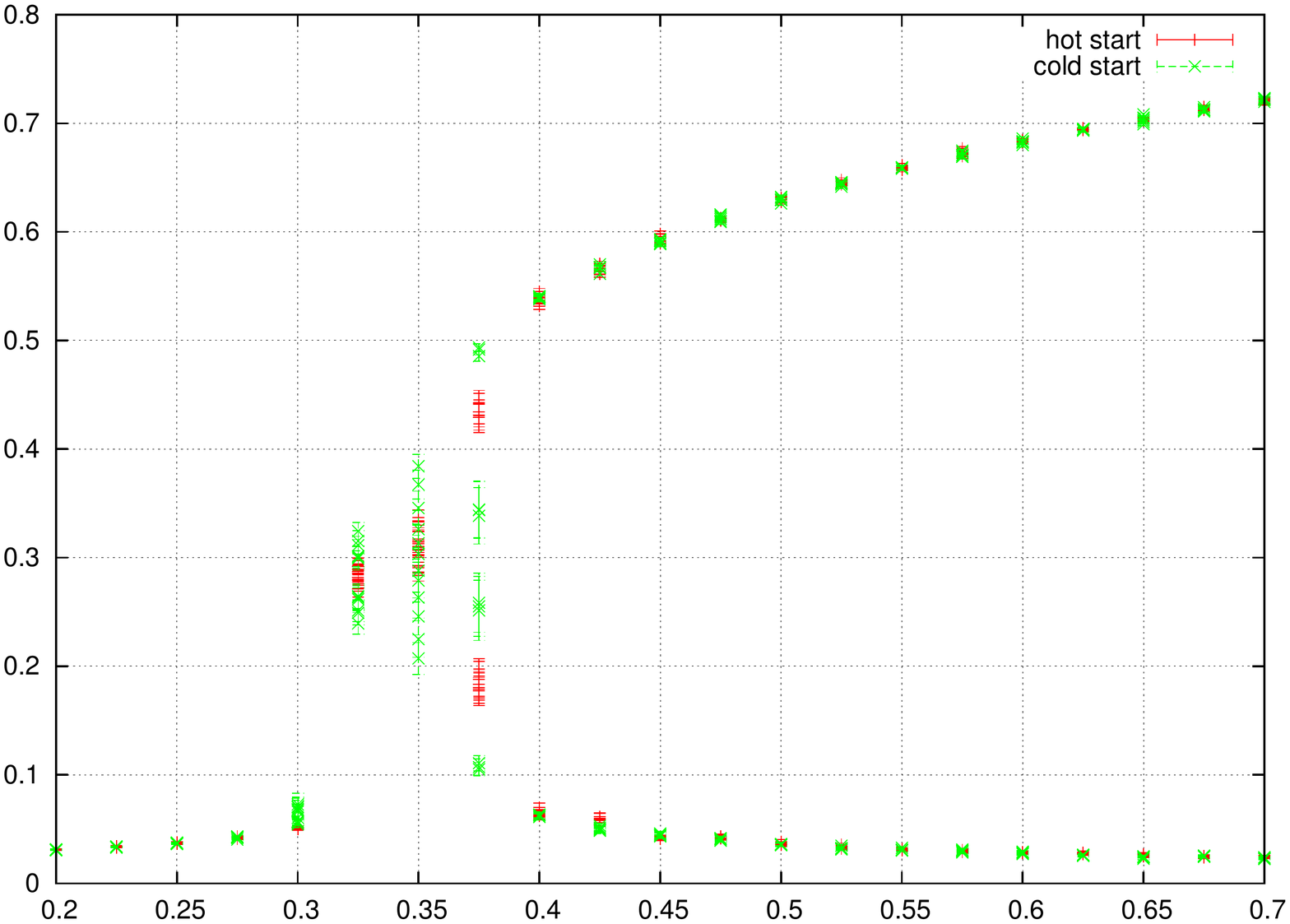}
\caption{The values of all twelve $\<|M_{\mu,\nu}|\>$ plotted
versus $b$ for $N=20$ (upper panel) and $N=40$ (lower panel),
using $\rho_{\rm clock}$ and self-averaging.}
\label{hyst_map_M}
\end{figure}

Below, and in the next three sub-sections, we show that this impression is
wrong. A clear signal for this can be seen in Fig.~(\ref{hyst_map_M}),
which shows how the absolute values $|M_{\mu,\nu}|$ depend on $b$.
We recall that the $M_{\mu,\nu}$ transform non-trivially under
center and reflection symmetries.
The discussion of Sec.~\ref{implication_of_locking} implies
that, for reduction to hold, 
the $|M_{\mu,\nu}|^2$ must have expectation values of $O(1/N)$, 
and thus that the $\langle |M_{\mu,\nu}|\rangle$ 
should fall to zero as $N\to\infty$.
In fact what we find is that some of the $M_{\mu,\nu}$ fluctuate
around $O(1)$ values in the weak-coupling phase.
This is the first indication that reduction does not hold
in the QEK model. 

This result calls for a more detailed study of systematic
errors. These include the possibility of very large $O(1/N)$ corrections
(i.e. that the non-zero values for $\langle |M_{\mu,\nu}|\rangle$ would
decrease for large enough $N$),
dependence on the choice $\rho(p)$ or on the self-averaging strategy, 
and the possibility that the simulations did not,
in fact, equilibrate (and that given enough updates would tunnel into a
``vacuum'' that satisfies reduction).
In addition, a more accurate determination of the
transition coupling $b_t$ would allow a direct test of reduction.
This is the coupling at which
the QEK model goes through a first order transition, 
and, if reduction holds, should equal the bulk-transition
coupling $b_{\rm bulk}$ of the infinite-volume, large-$N$, pure-gauge theory.
In the past,
the numerical proximity of $b_t$ and $b_{\rm bulk}$ was considered 
to be evidence in favor of large-$N$ 
equivalence~\cite{BHN2,Okawa2,BM,Lewis,Carlson}, but the
calculations of $b_t$ were not of high accuracy. 
In the next sub-sections
we attempt to address all these issues.

\subsection{Precise measurements of the plaquette}
\label{more_precise_plaquette}

In this section we perform high-precision measurements of 
$u_p$ for $b=0.4$, $0.45$ and $0.5$, values chosen to allow
comparison with results from the large volume simulations
of Ref.~\cite{TV,Mike}. We use and test both strategy A (explicit quenched
averaging), and strategy B (self-averaging), and in addition study
different choices for the measure $\rho(p)$. 
To implement strategy A we draw a new choice for $p$ 
(drawn randomly with weighting $\rho(p)$)
after a fixed number of equilibration and measurement sweeps.
The simulation parameters are given in Table~\ref{more_precise_tab1}.
For strategy B we simply use very long runs with a fixed 
choice of $p$.
Details are given in Table~\ref{more_precise_tab2}.

\begin{table}
\setlength{\tabcolsep}{4mm}
\begin{tabular}{ccccc}
\hline\hline  $N$ & choices of $p$ & equilibration updates & measurements & $\rho(p)$
\\  \hline 
$20$, $40$, $80$ & $20$ & $1000$ & $5000$ & clock \\ 
$40$ & 20 & $5000$ & $1000$ and $2000$ & uniform \\\hline 
\hline
\end{tabular}
\caption{Simulation details for measurements of $u_p$ using strategy A---explicit
quenched averaging. The number of equilibration updates and measurements is
for each value of $p$. We use $b=0.4$, $0.45$ and $0.5$ in all cases.
}
\label{more_precise_tab1}
\end{table}

\begin{table}
\setlength{\tabcolsep}{4mm}
\begin{tabular}{cccc}
\hline\hline  $N$
& equilibration updates & measurements & choice of $\rho(p)$ and $p$ 
\\\hline\hline
$20$, $40$, $80$ & $5000$ & $100000$ 
& \parbox{4cm}{clock ($p=p_{\rm locked}$ \\ or permutation)} 
\\ 
$20$, $40$ &$5000$ & $100000$ & uniform  \\ 
$80$ &  $1000-5000$ & $10000$ & uniform  \\ 
$50$ & $1000$ & $20000$ & clock ($p=p_{\rm locked}$) \\ 
$100,125,150,200$& $500$ & $1000$ & clock ($p=p_{\rm locked}$) \\ 
$16$, $81$ & $5000$ & $100000$ & BZ \\ 
\hline\hline
\end{tabular}
\caption{Parameters of simulations used to calculate $u_p$ 
using strategy B---self-averaging.
We use $b=0.4$, $0.45$ and $0.5$ except for $N=50$, where we only use
$b=0.4$ and $0.5$, and for $N=100$, $125$, $150$, $200$, where we only use
$b=0.4$.
}
\label{more_precise_tab2}
\end{table}

\begin{figure}[htb]
\centerline{
\includegraphics[angle=-90,width=10cm]{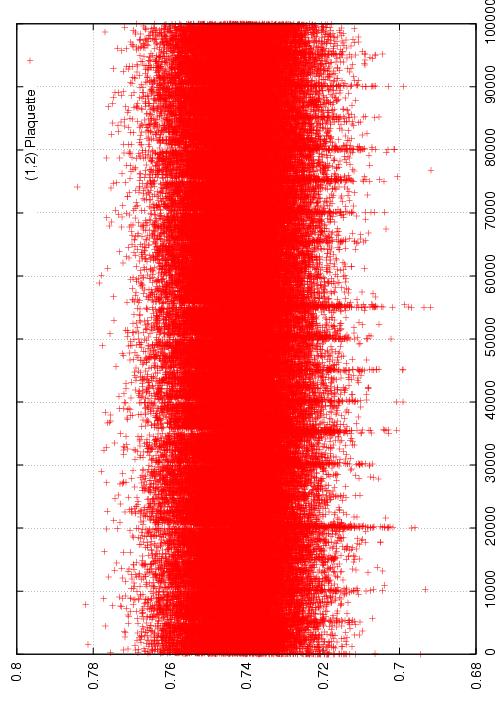}
\includegraphics[angle=-90,width=10cm]{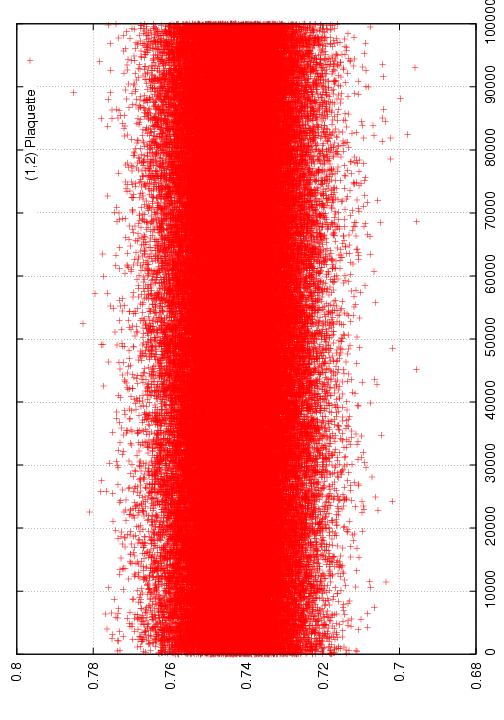}
}
\vskip 0.5cm
\centerline{
\includegraphics[angle=-90,width=10cm]{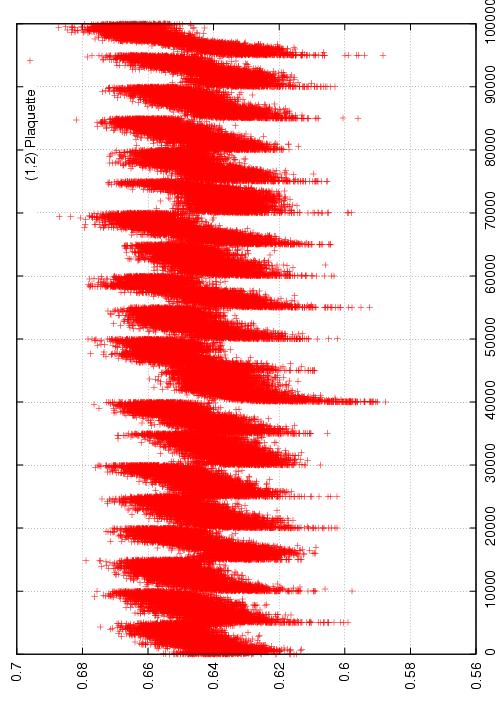}
\includegraphics[angle=-90,width=10cm]{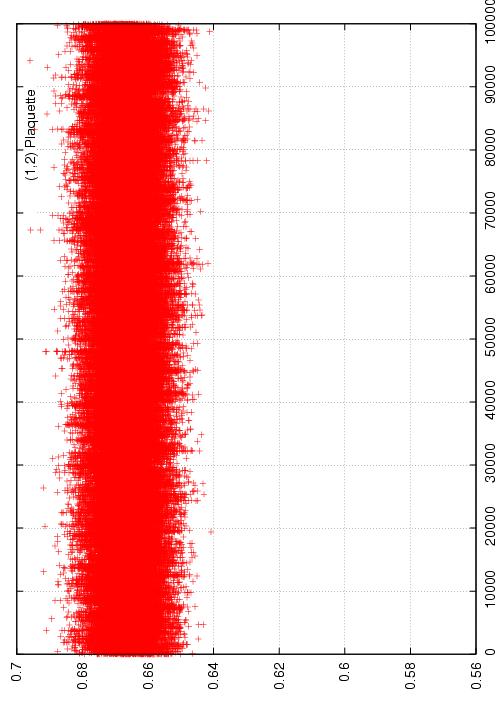}
}
\caption{The ($\mu=1$, $\nu=2$) plaquette variable vs measurement number. Equilibration updates
are not shown. The upper two panels are for 
$SU(40)$ with $b=0.5$, the lower two for $SU(80)$ with $b=0.4$.
The left-hand plots are for strategy A, in which a random permutation
of the clock momenta is generated every 5000 measurements.
The right-hand plots are for strategy B, with $p=p_{\rm locked}$.
}
\label{MC_time_plaq}
\end{figure}

We begin by comparing the plaquette time histories for the two
strategies. In Fig.~\ref{MC_time_plaq} we show results for $N=40$ at
$b=0.40$ and for $N=80$ at $b=0.50$, in both cases using $\rho_{\rm clock}(p)$. 

\begin{figure}[htb]
\centerline{
\includegraphics[angle=-90,width=15cm]{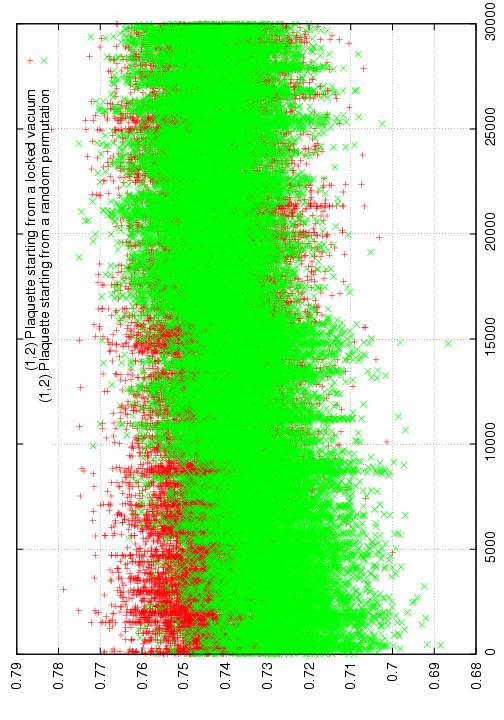}
}
\caption{The $(\mu,\nu)=(1,2)$ plaquette variable for $SU(40)$ at $b=0.50$ vs. 
update with strategy B (self-averaging) with $\rho_{\rm clock}$. 
[Red] plusses are for $p=p_{\rm locked}$,
while [green] crosses are for a random permutation of the clock momenta.}
\label{MC_time_plaq_random}
\end{figure}

The results for $N=40$ suggest that, once equilibrated, both strategies
give similar results. The ``tails'' below the main band for strategy A indicate,
however, that insufficient equilibration sweeps were included.
This effect is much clearer for $N=80$, because of the smaller fluctuations.
To see what happens with sufficient
equilibration, we present in Fig.~\ref{MC_time_plaq_random} the time
histories for $SU(40)$ at $b=0.50$, obtained using strategy B, with both
$p=p_{\rm locked}$ and a single fixed random permutation.
After a long period in a metastable state, with a relatively
low value of $u_p$, the system does appear to tunnel into a state with
a plaquette value consistent with that for $p=p_{\rm locked}$.
The time this requires (about 15000 measurements) exceeds, however, that used in our strategy A runs.

We tentatively conclude that self-averaging 
works at least approximately for the plaquette, 
given long enough equilibration times.
This conclusion is supported by the results from the other simulations
listed in Tables~\ref{more_precise_tab1} and \ref{more_precise_tab2}.
To illustrate this, we collect,
in Tables~\ref{summary_comp_strategies_plaq}
and \ref{more_precise_plaq2}, the average values of $u_p$ 
using both strategies and for clock and uniform densities.
For strategy B we show only results from runs that were equilibrated.
For strategy A all results are suspect because of the
thermalization issues discussed above and illustrated by
Fig.~\ref{MC_time_plaq}. We nevertheless include them as a comparison.
The table also includes the best estimates for the plaquette
values for $SU(\infty)$, obtained by extrapolating from large-volume
simulations \cite{Mike-Helvio}. These are the numbers that would be reproduced by
a large $N$ extrapolation of QEK results were reduction to hold.

We first comment on the results using strategy B.
We first note (from Table~\ref{summary_comp_strategies_plaq})
that, in all cases, the final plaquette
is independent of the choice of input momenta for $\rho=\rho_{\rm clock}$.
This is the expected self-averaging for a center-invariant quantity.
More striking is that, as $N$ increases, results from strategy B
using $\rho_{\rm uniform}$ appear to converge to those from $\rho_{\rm clock}$.
This gives us confidence that we are not observing systematic errors due
to the choice of $\rho$, and that the systematic differences with
the results from strategy A are due to lack of equilibration of the latter.

\begin{table}
\setlength{\tabcolsep}{4mm}
\begin{tabular}{clccc}
\hline\hline
$b$,  $u_p$ for $SU(\infty)$
& Strategy and $\rho(p)$ & $SU(20)$ & $SU(40)$ & $SU(80)$ \\ \hline
\multirow{5}{*}{$b=0.50$, $u_p\simeq 0.7182$} 
 & A, clock$^\star$ & 0.7294(2) & 0.7256(2) & 0.7223(4)
 \\ & B, clock ($p=p_{\rm locked}$) & 0.7396(5) & 0.7425(2)
 & 0.7429(1) \\ & B, clock ($p={\rm random}$) & 0.739(1)
 & 0.7424(1) & 0.7429(1) \\ & A, uniform$^\star$ &
 0.7401(3) & -- & --\\ & B, uniform & 0.7483(3) &
 0.7405(2) & 0.7432(1) \\ \hline
\multirow{5}{*}{$b=0.45$, $u_p\simeq 0.6795$} 
  & A, clock$^\star$ & 0.6968(5) & 0.7090(1) &
  0.6867(8) \\ & B, clock ($p=p_{\rm locked}$) & 0.7035(9)
  & 0.7094(1) & 0.7100(1) \\ & B, clock ($p={\rm random}$)
  & 0.703(1) & 0.7095(1) & 0.7101(1) \\ & A,
  uniform$^\star$ & 0.7082(4)& -- & --\\ & B, uniform &
  0.7153(2) & 0.7086(2) & 0.7100(1)\\ \hline
\multirow{5}{*}{$b=0.40$, $u_p\simeq 0.6259$} 
  & A, clock$^\star$ & 0.6533(5) & 0.6489(7) &
  0.645(1) \\ & B, clock ($p=p_{\rm locked}$) & 0.66014(54)
  & 0.6651(2) & 0.6665(1) \\ & B, clock ($p={\rm random}$)
  & 0.6595(5) & 0.6645(3) & 0.6665(1) \\ & A,
  uniform$^\star$ & 0.6648(5) & -- & --\\ &B, uniform &
  0.6737(5) & 0.6652(2) & 0.6642(3)\\ \hline\hline
\end{tabular}
\caption{Comparison of plaquette expectation values 
between averaging strategies
and different choices of $\rho(p)$. 
The results from strategy A are
denoted by a ``$^\star$'' to indicate that they are suspect
due to a possible lack of equilibration (see text). 
The first column includes the estimates
for $SU(\infty)$ based on extrapolations using large-volume simulations from Ref.~\cite{Mike-Helvio}.}
\label{summary_comp_strategies_plaq}
\end{table}

\begin{table}
\setlength{\tabcolsep}{4mm}
\begin{tabular}{ccccc}
\hline \hline
$SU(50)$ & $SU(100)$ & $SU(125)$ & $SU(150)$ & $SU(200)$ \\ \hline
0.6662(9)  & 0.6647(3) & 0.6658(3) & 0.6667(3) & 0.6670(2) \\
\hline\hline
\end{tabular}
\caption{Additional results for $u_p$ obtained with strategy B 
at $b=0.40$ with $\rho_{\rm clock}$ and $p=p_{\rm locked}$. 
}
\label{more_precise_plaq2}
\end{table}

\begin{figure}[htb]
\centerline{
\includegraphics[width=11cm]{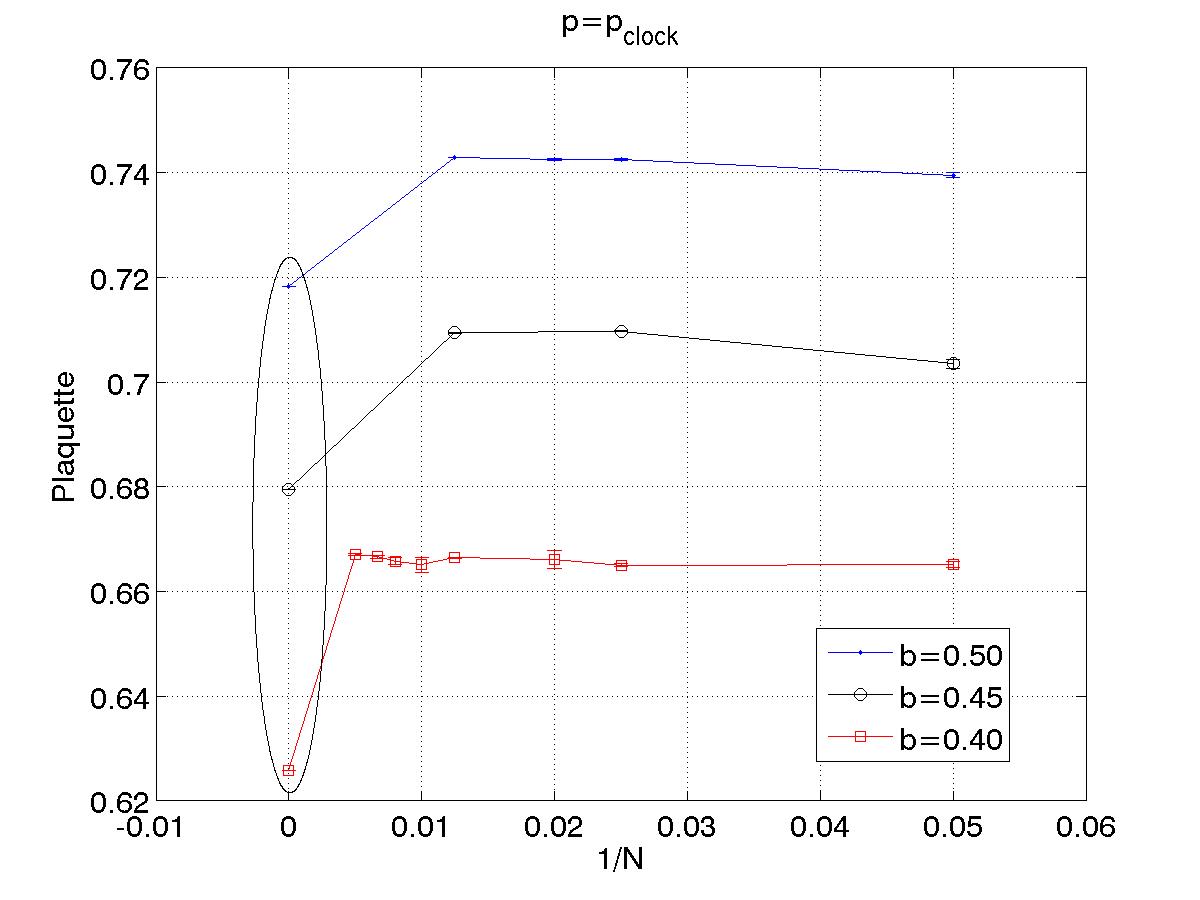}
}
\caption{$u_p$ versus $1/N$, for $b=0.4$, $0.45$ and $0.5$, compared to
the expectations for infinite-volume $SU(\infty)$ gauge theory (presented inside
the ellipse). Lines are only to guide the eye.
Results shown are for $\rho=\rho_{\rm clock}$ with $p=p_{\rm locked}$. 
}
\label{plaq2largeN}
\end{figure}

The most important comparison is with the results for the
infinite-volume $SU(\infty)$ theory. To make this more precise
, we extended the results at $b=0.4$ up to $N=200$ (see Table~\ref{more_precise_plaq2}).
The resulting comparisons are shown in Fig.~(\ref{plaq2largeN}).
We have plotted $u_p$ versus $1/N$, since this is the expected $N$ dependence
in the QEK model.
Our results show a fairly smooth extrapolation to $N=\infty$, with
small corrections whose dependence on $1/N$ we cannot definitely
determine. 
We do not perform a detailed fit, however, since it is clear that,
regardless of the precise form of the subleading terms,
our results extrapolate to significantly higher values of $u_p$
than those of the infinite-volume lattice gauge theory.
This discrepancy clearly shows that the QEK model does
not reproduce  the physics of the large-$N$ gauge theory.

\begin{figure}[htb]
\centerline{
\includegraphics[width=11cm]{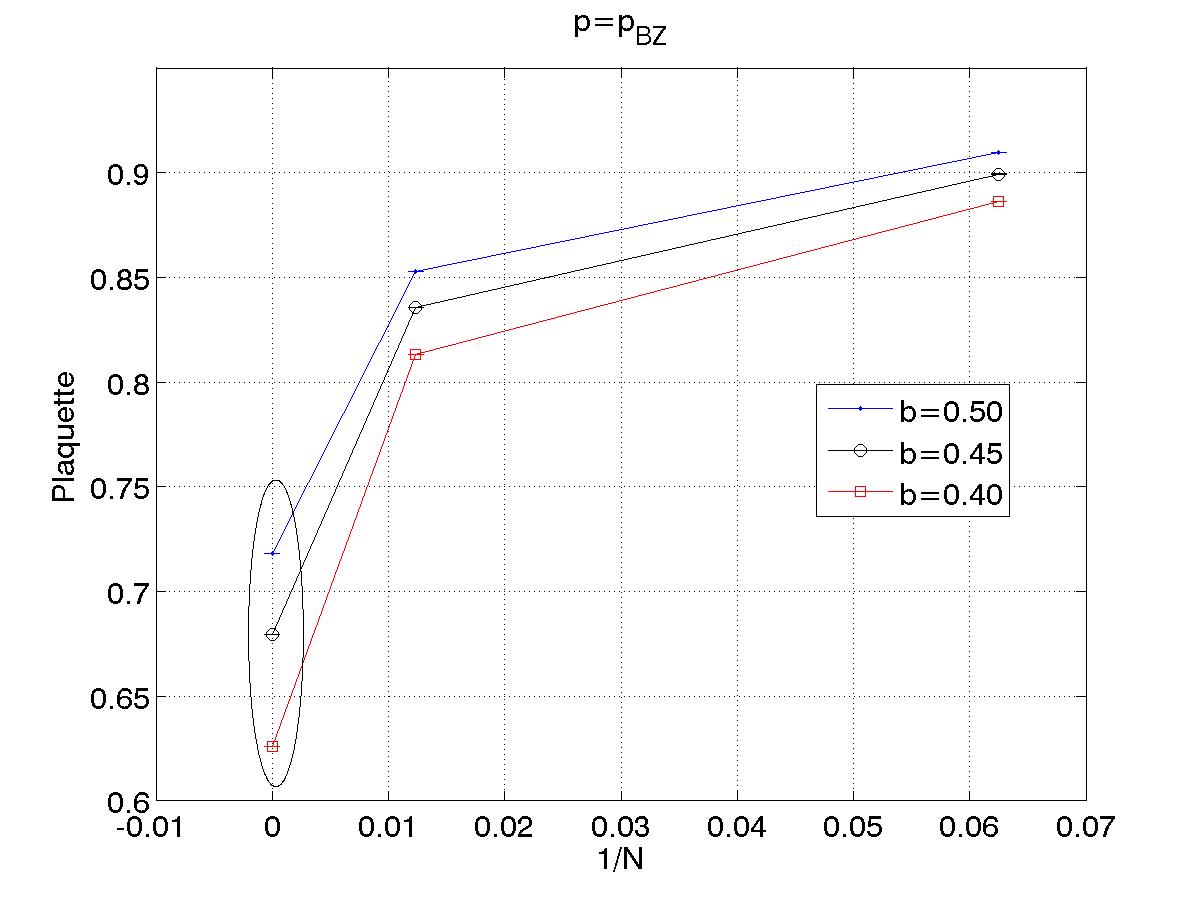}
}
\caption{As in Fig.~\protect\ref{plaq2largeN}, but
for $\rho_{\rm BZ}$.
Note that the vertical scale differs from Fig.\protect\ref{plaq2largeN}.
}
\label{plaqBZlargeN}
\end{figure}

We have also obtained results using $\rho_{\rm BZ}$. These are
collected in Table~\ref{more_precise_plaq1}, 
and the comparison to the lattice large-$N$ result is shown in Fig.~\ref{plaqBZlargeN}.
The discrepancy with infinite-volume $SU(\infty)$ values is significantly
larger in this case, a point we return to below.

\begin{table}
\setlength{\tabcolsep}{4mm}
\begin{tabular}{cccc}
\hline \hline
$N$ & $b=0.4$ & $b=0.45$ & $b=0.5$ \\ \hline 
$16$ & 0.88627(5) & 0.89935(4) & 0.90961(3)  \\ 
$81$ & 0.812323(2) & 0.83546(1) & 0.85291(1)  \\ 
\hline\hline
\end{tabular}
\caption{Results for $u_p$ using strategy B for $\rho_{\rm BZ}$.}
\label{more_precise_plaq1}
\end{table}

\subsection{Precise measurements of the $M_{\mu,\nu}$}
\label{more_precise_M}

\begin{figure}[thb]
\centerline{
\includegraphics[angle=-90,width=11cm]{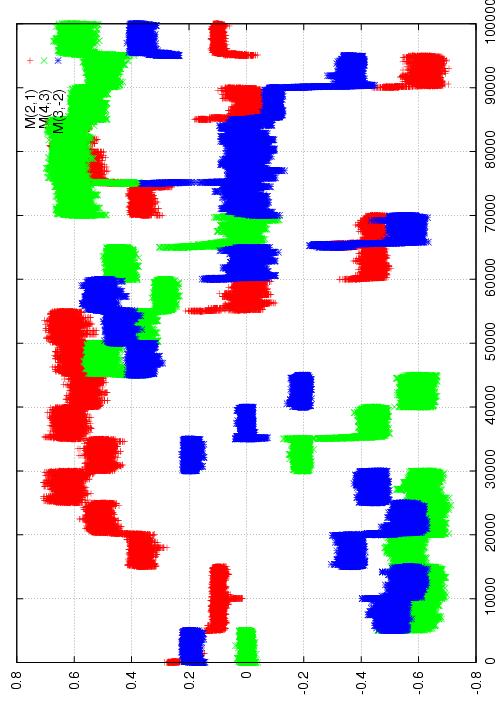}
}
\caption{Real part of $M_{\mu,\nu}$ versus measurement number for
$N=40$ and $b=0.50$ with $\rho_{\rm clock}(p)$. 
The figure shows a sequence of $20$ Monte-Carlo runs, 
each with $1000$ equilibration updates (not shown) followed by
$5000$ measurements, and each with a randomly chosen permutation
of the clock momenta. For clarity, we only present $M_{2,1}$
([red] plusses), $M_{4,3}$ ([green] crosses), 
and $M_{3,-2}$ ([blue] fancy crosses).
}
\label{history_M_clock}
\end{figure}
\begin{figure}[bht]
\centerline{
\includegraphics[angle=-90,width=11cm]{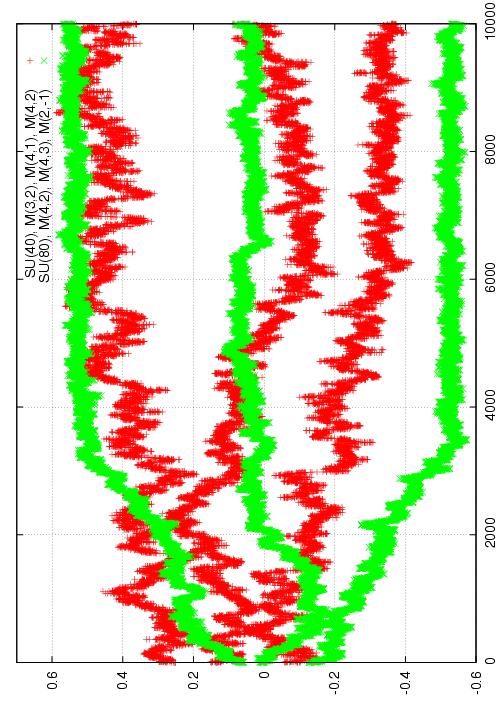}
}
\caption{Real part of $M_{\mu,\nu}$ versus measurement number for $\rho_{\rm uniform}(p)$,
with a single random $p$, at $b=0.5$ and for $N=40$ ([red] plusses)
and $80$ ([green] crosses).
We show only  $M_{3,2}$, $M_{4,1}$, and $M_{4,2}$ for $SU(40)$, 
and $M_{4,2}$, $M_{4,3}$, and $M_{2,-1}$ for $SU(80)$.}
\label{history_M_uniform}
\end{figure}

To elucidate the nature of the breakdown of reduction, we present here
results for the ``order parameters'' $M_{\mu,\nu}$. We use the
same simulation parameters as in the previous section.
We recall that, for reduction to hold, 
$\langle |M_{\mu,\nu}|^2\rangle_p$ should be no larger
than $O(1/N)$ for all $\mu$ and $\nu$.
Furthermore, for the case of $\rho_{\rm clock}$, the expectation
values $\langle M_{\mu\nu}\rangle_p$ are true order parameters for
spontaneous breakdown of the center symmetry.

We begin by presenting, in Fig.~\ref{history_M_clock},
the Monte-Carlo time history of the real parts
of a selection of the $M_{\mu,\nu}$,
using $\rho(p)=\rho_{\rm clock}(p)$ and strategy A
(explicit quenched averaging),
for $N=40$ and $b=0.50$.
This is the run for which we have previously 
shown the plaquette in the upper-left panel of
Fig.~\ref{MC_time_plaq}.
We clearly see equilibration into distinct ``vacua'' for
different choices of input momenta, and in several cases
we can see the tail-end of what appears to be a tunneling
process. The values of ${\rm Re}(M_{\mu,\nu})$ either oscillate
around zero or around values of $O(1)$. The latter indicate
SSB of the center symmetry, and the 
presence of the locked momenta discussed in Sec.~\ref{sec:locking}.

We show a similar plot for $\rho_{\rm uniform}(p)$
in Fig.~\ref{history_M_uniform}, except that
we use only a single random choice of $p$, 
have longer runs to assure equilibration,
and present results for both $N=40$ and $80$.
After a long equilibration period, the $M_{\mu,\nu}$ 
at both $N$ fluctuate around what we assume to be 
vacuum values. We note that
the fluctuations are smaller for the larger $N$, as expected in general. 
We see this behavior throughout our study.
The crucial observations, however, are that some of the $M_{\mu,\nu}$
fluctuate around non-zero $O(1)$ values (indicating locked momenta),
and that these values are comparable for both $N$.
This implies that the left-hand side of \Eq{Mloop1} is of $O(1)$,
and reduction does not hold.

\begin{figure}[htb]
\centerline{
\includegraphics[angle=-90,width=9cm]{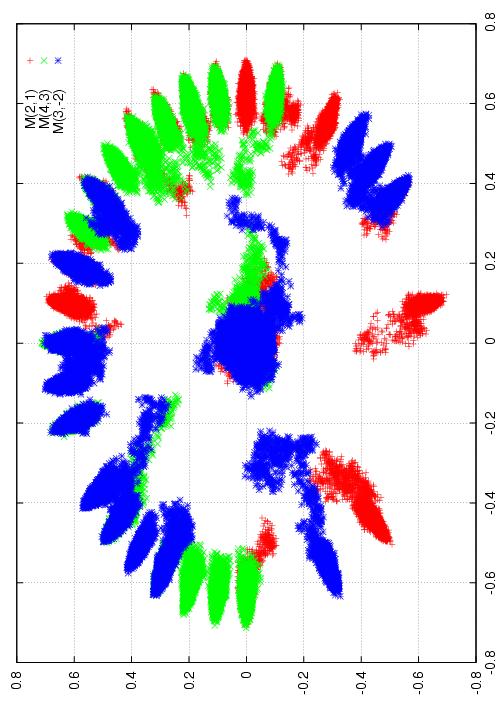}
\includegraphics[angle=-90,width=9cm]{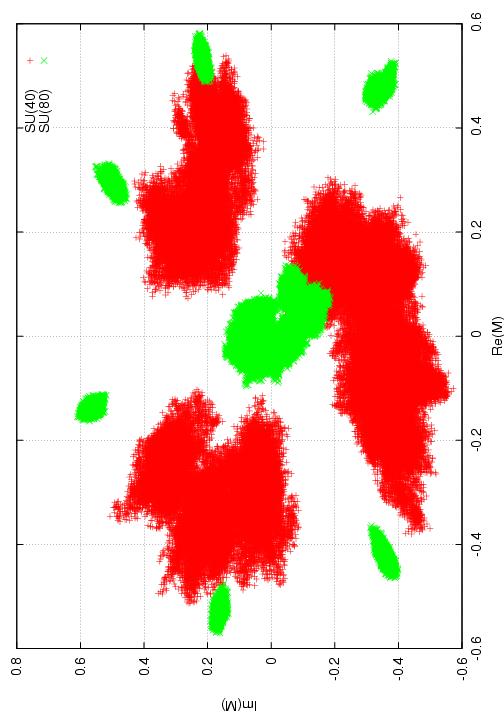}
}
\caption{
{Left panel:} scatter plot of the data that appears in the Fig.~\ref{history_M_clock}. 
{Right panel:} scatter plot of all twelve $M_{\mu,\nu}$ obtained 
from runs with $\rho(p)=\rho_{\rm uniform}(p)$ for $SU(40)$ ([red] plusses) 
and $SU(80)$ ([green] crosses) at $b=0.50$, with only equilibrated
results shown.
\label{scatter_M_A}}
\end{figure}

It is also instructive to look at the full complex values of the
$M_{\mu,\nu}$. 
In the left panel of Fig.~\ref{scatter_M_A} we show the scatter plot
for the same data-set used in Fig.~\ref{history_M_clock}.
Apart from ``equilibration tails'', we see that the simulations settle
down into vacua in which a given $M_{\mu,\nu}$ either fluctuates around $0$
or around $m_0 \exp(2 \pi i n/40)$, with $n$ an integer and $m_0\approx 0.65$.
The different vacua are related by (an appropriate subgroup of
the $(Z_N)^4$) transformations.
This is qualitatively consistent with what we would expect with locked vacua
when fluctuations are included. Without fluctuations,
the locked vacua have half of the $M_{\mu,\nu}$ vanishing, and the other half
of the form $\exp(2 \pi i n/N)$. The fluctuations reduce the magnitude
from unity to $m_0$. Note that this reduction is greater than one would predict
from a simple mean link model, in which $m_0 \approx \sqrt{u_p} \approx 0.86$.
This may be a consequence of the fact that the partial unlocking of momenta
can reduce $|M_{\mu\nu}|$ while leaving the plaquette unchanged.

This figure gives a very clear illustration of the way in which
$\langle M_{\mu,\nu}\rangle_{\rm QEK}$ vanishes when using $\rho_{\rm clock}(p)$.
One is instructed to average over input momenta which are permutations
of the clock momenta. For given input momenta, the dynamics picks
a (partially) locked vacuum. As the average is taken, each
$M_{\mu,\nu}$ will end up with equal probability in the center near the origin,
or in the ``ring'' of radius $m_0$, and in the latter case with equal probability
in each of the $N$ vacua. In this way $M_{\mu,\nu}$ will average to zero.
As noted in Sec.~\ref{analytics}, the dynamics will determine whether,
for a given input momenta and as $N\to\infty$, the theory gets trapped in
a single vacuum or moves between them. Our numerical results strongly
indicate the former, in which case (for $\rho_{\rm clock}$) SSB 
is occurring.

A similar scatter plot for $\rho_{\rm uniform}$ is shown in
the right panel of Fig.~\ref{scatter_M_A}. In this case all twelve
$M_{\mu,\nu}$ are shown for each $N$ (not just the three for each $N$ shown in
Fig.~~\ref{history_M_uniform}), and we display only measurements
after equilibration.
For $\rho_{\rm uniform}$ there is no center symmetry, but we do see (most
clearly for $N=80$) the expected pattern for locked momenta
of six non-zero and six near-zero magnitudes.
(Note that some of the [red] $N=40$ points near the origin are obscured
by the [green] $N=80$ points.)
We also observe no reduction in the $O(1)$ magnitudes as $N$ increases
from $40$ to $80$---indeed the magnitudes seem to increase. 
This we take as strong evidence for the breakdown of reduction.

\bigskip

\begin{figure}[htb]
\centerline{
\includegraphics[angle=-90,width=9cm]{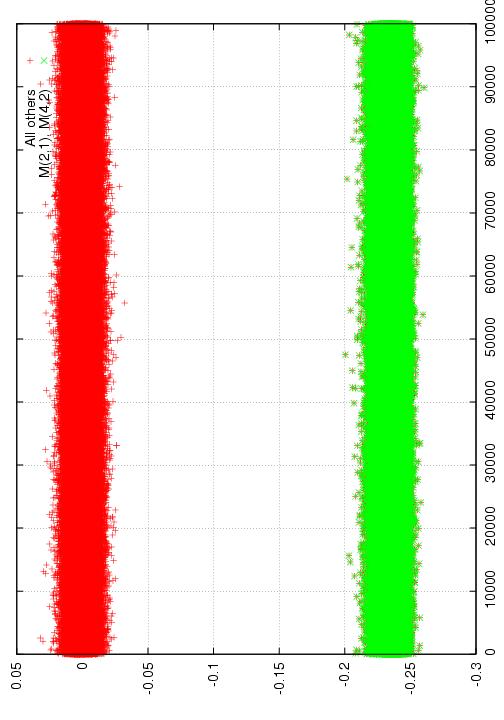}
\includegraphics[angle=-90,width=9cm]{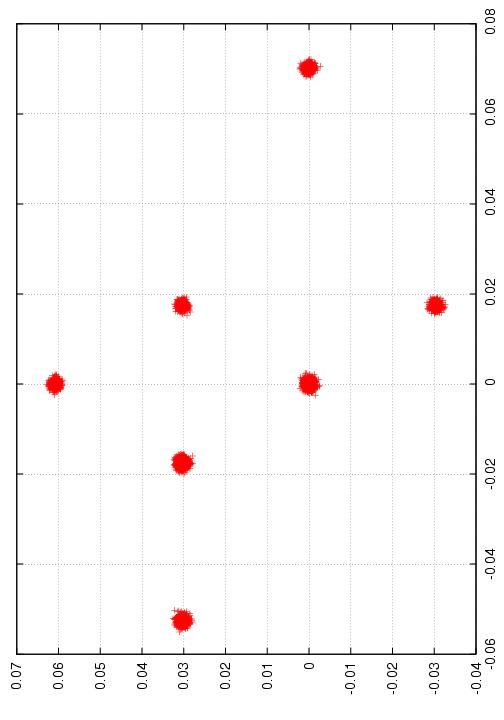}
}
\caption{Results for $M_{\mu,\nu}$ with $\rho_{\rm BZ}$.
{Left panel:}  Time history (versus measurement number) of all six
independent $M_{\mu,\nu}$ for $N=16$ and $b=0.40$. 
{Right panel:} Scatter plot of all $M_{\mu,\nu}$ 
for $N=81$ and $b=0.70$.}
\label{Bars_stuff}
\end{figure}

Finally, we consider $\rho_{\rm BZ}$. We show results
obtained only from a hot start.\footnote{%
The fluctuations in the runs beginnings from cold starts were too small
to allow  the simulation to forgets its initial state, be it a state with
zero or nonzero $M_{\mu,\nu}$}
In the left panel of Fig.~\ref{Bars_stuff},
we show the time history of all the $M_{\mu,\nu}$ for $N=16$ and $b=0.40$.
Recalling the definition of $\rho_{\rm BZ}$ from
\Eq{Bars_p}, we note that, since $K=\sqrt[4]{16}=2$,
all $p^a_\mu$ are either $0$ or $\pi$. This means that
the $M_{\mu,\nu}$ are real, and that $M_{\mu,\nu}=M_{\mu,-\nu}$ 
(so there are only 6 independent $M_{\mu,\nu}$).
Furthermore, the center symmetry is only $(Z_2)^4$, 
although this symmetry group is still
sufficient to forbid expectation values for the $M_{\mu,\nu}$.
What we see from the figure is that while
four of the $M_{\mu,\nu}$ fluctuate around zero, two of them 
($M_{1,2}$ and $M_{4,2}$ )
acquire nonzero expectation values that break the $(Z_2)^4$ symmetry.

We can understand this pattern of expectation values in the
following way. The input momenta [defined in \Eq{Bars_p}] are such that
\begin{eqnarray}
\Lambda_1 &=& {\rm diag}(\sigma_3,\sigma_3,\dots)\,,\ \
\Lambda_2 = {\rm diag}(\bm{1}_2,-\bm{1}_2,\bm{1}_2,-\bm{1}_2,\dots)\,,\ \
\nonumber\\
\Lambda_3 &=& {\rm diag}(\bm{1}_4,-\bm{1}_4,\bm{1}_4,-\bm{1}_4)\,,\ \
\Lambda_4 = {\rm diag}(\bm{1}_8,-\bm{1}_8)\,,
\end{eqnarray}
where $\bm{1}_n$ indicates (the diagonal part of) an $n$-dimensional
identity matrix. With these matrices, and assuming $U_\mu \approx \Lambda_\mu$
(i.e. ignoring fluctuations due to the $V_\mu$), all the $M_{\mu,\nu}$ vanish.
By a single transposition, however, one can change $\Lambda_2$ to
\begin{equation}
\Lambda'_2 = {\rm diag}
(-\sigma_3,-\bm{1}_2,\bm{1}_2,-\bm{1}_2,\bm{1}_2,-\sigma_3,\bm{1}_2,-\bm{1}_2)\,.
\end{equation}
One then finds, in the same approximation of ignoring fluctuations,
that there are two non-zero $M_{\mu,\nu}$: $M_{1,2}=M_{2,4}=-0.25$. 
Fluctuations will reduce the average from this value.
Thus this scenario provides a possible explanation for 
the results of the left panel of Fig.~\ref{Bars_stuff}.
This is, in fact, one of many choices of transpositions that
leads to this pattern of expectation values.
Furthermore, all the patterns of values for the $M_{\mu,\nu}$ that
we have observed in our $N=16$ runs can be explained similarly.
 
\begin{figure}[!htb]
\centerline{
\includegraphics[trim=0.5cm 6cm 1cm 6cm, clip=true,width=12cm]
{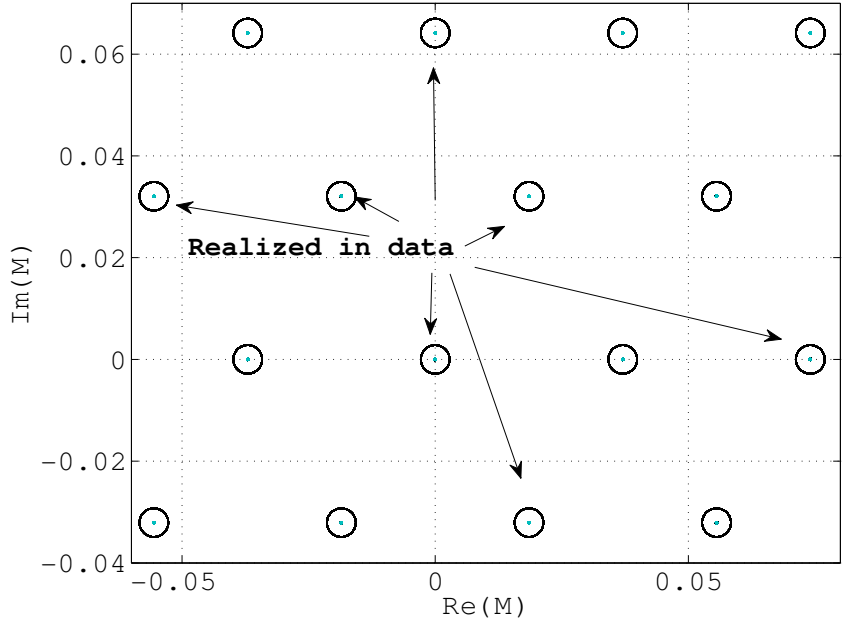}}
\caption{Some of the possible values for $M_{\mu,\nu}$ 
in the $N=81$ case, assuming no fluctuations, 
i.e. when the $p_\mu^a$ are permutations of $p_{BZ}$. 
Note that, in a given simulation, one expects only some 
(at most 12) of these values to be realized.
This figure should be compared 
to the right panel of Fig.~\ref{Bars_stuff}.
\label{Bars_su81_analysis}}
\end{figure}

We see an analogous phenomenon for $N=81$. Here, since $K=3$,
the center symmetry is $(Z_3)^4$. The right panel of Fig.~\ref{Bars_stuff}
shows a scatter plot of all the (now twelve) $M_{\mu,\nu}$
from a simulation at $b=0.70$. One can
understand this figure by calculating the possible
values of $M_{\mu,\nu}$ that are obtained by permuting the
elements of the initial $\Lambda_\mu$, and ignoring fluctuations.
The result is shown in Fig.~\ref{Bars_su81_analysis}, and 
is clearly a good description of
what we see in the right panel of Fig.~\ref{Bars_stuff}.

We note that, unlike for $\rho_{\rm clock}$, 
the BZ weight function does not lead to
complete or nearly-complete momentum locking. 
In a completely
locked state, all the $\Lambda_\mu$ are equal up to center
and reflection transformations, and this leads, in the example of $N=2$, 
to all six independent $M_{\mu,\nu}$ being close to $\pm 1$.
To reach such a locked state requires many transpositions, however,
and our results suggest that only a few transpositions have occurred.

\bigskip

In summary, the numerical results presented in this sub-section 
indicate that some of the ``order parameters'' $M_{\mu\nu}$ acquire
$O(1)$ expectation values, which, as described in Sec.~\ref{analytics},
is inconsistent with large-$N$ reduction for the QEK model.
For the weight functions $\rho_{\rm clock}$ and $\rho_{\rm BZ}$, 
the expectation values for the $M_{\mu,\nu}$ spontaneously break the
center and reflection symmetries.\footnote{%
The breaking pattern depends on the extent of locking.
For complete locking, and $\rho_{\rm clock}$,
the breaking is $Z_N^4\to Z_N$,
where the remaining symmetry is the diagonal $Z_N$.}
This breakdown is not apparent in the simplest open loops,
i.e. $\<\tr U_\mu^n\>$ with $n<N$, but is exhibited by more
complicated objects like the ``corner'' variables $M_{\mu,\nu}$. 
For the uniform and clock distributions,
the actual values of the $M_{\mu,\nu}$ are qualitatively consistent with
the ``momentum-locking'' predicted by the weak-coupling analysis.
That analysis, however, could not determine whether the symmetry-breaking
or cluster-decomposition-violating scenario would hold.
Our numerical results clearly favor the former.

\subsection{Precise measurements of the transition coupling $b_t$}
\label{bt_w_WL}

The plaquette data in Fig.~(\ref{hyst_map}) strongly suggest that the
QEK model has a first order phase transition for $b$ somewhere
in the range $0.30-0.33$. 
This was already noted in the early QEK literature~\cite{BHN2,Okawa2,Bhanot,BM}, 
and the transition was assumed to be the same as
that which occurs in the $SU(\infty)$ gauge theory at 
$b_{\rm Bulk}\simeq 0.36$ (the ``bulk transition''). 
The $\sim 10\%$ discrepancy was attributed to $O(1/N)$ corrections
and/or other systematic errors. In this section we
revisit this issue, and, in particular,
attempt to greatly reduce the systematic errors in the
determination of $b_t$.

The main source of uncertainty is the strongly first-order nature
of the transition, and the consequent metastability.
The strength of the transition is indicated by the size of the
jump in the plaquette, which is $\sim 0.3$.
Although, strictly speaking, there is no transition unless $N\to\infty$,
already for $N=50$ there is a significant hysteresis regime
of width $\Delta b \simeq 0.05$, and this width increases with $N$.
Thus an estimate of $b_t$ from Fig.~\ref{hyst_map} has an $O(15\%)$ error at
$N=50$, and this error too increases with $N$. 
It does not help to calculate $u_p$ on a denser grid, because
of the metastability. 

One way forward is to use re-weighting, making use of those
values of $b$ where tunneling between phases occurs. 
We expect the tunneling probability to fall exponentially
with $N^2$ (which counts the number of degrees of freedom and thus is
like the volume), and our results are qualitatively consistent with this.
We find that we can successfully use standard Ferrenberg-Swendsen (FS) 
re-weighting~\cite{FS} for $N=20$ and $30$, 
but for $N$ larger than about $40$ the method fails 
because tunneling ceases. 

To proceed we need a method which encourages tunneling.
We chose to
use the ``Wang-Landau'' re-weighting method,
developed recently in the field of statistical mechanics
\cite{WL}. This required adapting the method from spin-systems to
gauge theories, as well as developing a systematic way of estimating
errors. Presenting this analysis is beyond the scope of this paper,
and is presented in Ref.~\cite{WL_paper}.
We note only that this is an adaptive method of determining the density
of states, which includes a feature that forces motion through
configuration space.

\begin{table}
\setlength{\tabcolsep}{4mm}
\begin{tabular}{cllll}
\hline \hline
Type of re-weighting &$\ SU(20)$ & $\ SU(30)$ & $\ SU(40)$ & $\ SU(50)$ \\ \hline
Ferrenberg-Swendsen & 0.29598(5) & 0.30545(5) & -- & -- \\ 
Wang-Landau & 0.29544(37) & 0.30569(17) & 0.30968(20) & 0.31121(19) \\ \hline\hline
\end{tabular}
\caption{Values of the strong-weak transition
coupling $b_t$, obtained for $\rho_{\rm clock}(p)$
with $p=p_{\rm locked}$, with two different re-weighting methods.}
\label{bt_N_results}
\end{table}

We have carried out these calculations only for
$\rho=\rho_{\rm clock}$ and with input momenta being locked.
Since the algorithms are designed to ensure ergodicity, however, we
expect that the simulations will explore multiple permutations of the momenta,
i.e. will be self-averaging. Evidence in support of this expectation
is that we do see many tunnelings between the weak and strong phases for all $N$.
Table~\ref{bt_N_results} gives our results for the transition
coupling (defined as the peak in the susceptibility).
We find that the results from both techniques agree 
(when both are available)
despite the very small ($0.02\%-0.1\%$) statistical errors.

\begin{figure}[htb]
\centerline{
\includegraphics[width=20cm]{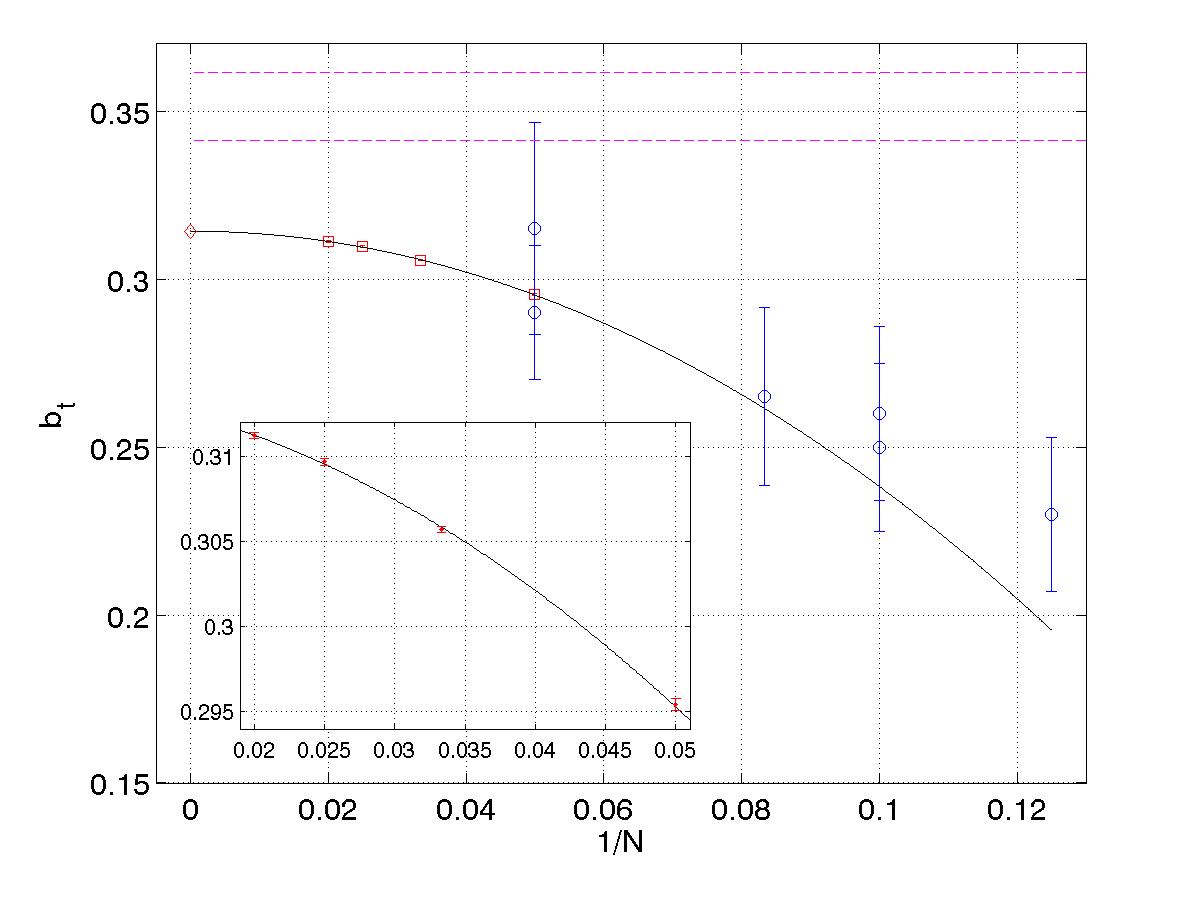}
}
\caption{The strong-to-weak transition coupling, $b_t$,
plotted versus $1/N$. [Red] squares show our results 
using the Wang-Landau algorithm from Table~\ref{bt_N_results}, while
[blue] circles are from Refs.~\cite{BHN2,Okawa2,Bhanot,BM}.
The solid black curve is the fit described in the text.
The [magenta] dashed horizontal lines give the range in which the bulk transition in $SU(12)$ should take place, 
according to hysteresis scans performed in a lattice theory \cite{Mike}.
The insert shows a close-up of our new data.}
\label{bt_vs_1_over_N}
\end{figure}

We plot these results versus $1/N$ in Fig.~(\ref{bt_vs_1_over_N}).
 For comparison, we include estimates of $b_t$ from the old numerical
 studies in Refs.~\cite{BHN2,Okawa2,Bhanot,BM}, as well as the most recent
 estimates of the coupling $b_{\rm bulk}$ at which the bulk transition
 occurs in the infinite volume $SU(\infty)$ gauge theory~\cite{Mike}. 
 While our new results are consistent with the
 old numerical studies of the QEK model, it is very unlikely that they
 extrapolate to the vicinity of $b_{\rm bulk}$. 
 We stress, however, that to make
 this observation, it is crucial to have very small errors,
 and this was  accomplished with the Wang-Landau algorithm.

We fit the Wang-Landau data (i.e. the second row in Table~\ref{bt_N_results})
to the form
\begin{equation}
b_t(N) = b_t(\infty) + \frac{A}{N} + \frac{B}{N^2}\,, 
\label{bt_N_fit}
\end{equation}
and find the fit parameters listed in Table~\ref{bt_N_fit_results}. 
The fits are of reasonable quality, and find values
for $b_t(\infty)$ which lie well below the estimate
$b_{\rm bulk}\simeq 0.36$. It is hard to quote a significance
for this discrepancy, since we do not have a good estimate
of the error in $b_{\rm bulk}$. If we use the error in our results,
the significance is between $\sim 50$ to $\sim 200$ $\sigma$.
We thus think it is very unlikely that
$b_t$ in the QEK model can be identified
with $b_{\rm bulk}$ of the
$SU(\infty)$ lattice gauge theory.

We have compared fits with and without the $1/N$ term, and find
that the former is slightly preferred, as shown in the Table.
It is this fit which is included in Fig.~(\ref{bt_vs_1_over_N}).
We have also attempted to fit simultaneously to the Wang-Landau results
for $N=40$ and $50$ and the (more accurate) FS results
for $N=20$ and $30$. This fit fails, quite likely because, given the
very high accuracy obtained with the FS method, we need to include
terms of $O(1/N^3)$.

\begin{table}
\setlength{\tabcolsep}{4mm}
\begin{tabular}{ccccc}
\hline\hline Type of fit & $b_t(\infty)$ & $A$ & $B$ & $\chi^2/{\rm d.o.f.}$\\ \hline
$A=0,B\neq 0$ & 0.3142(2) & -- & -7.59(18) & $1.45/1$ \\ 
$A\neq 0,B\neq 0$ & 0.3148(10) & -0.037(65) & -7.06(97) & 1.1/1 \\ 
\hline\hline
\end{tabular}
\caption{The parameters $b_t(\infty)$, $A$, and $B$, obtained 
from fitting the Wang-Landau data in Table~\ref{bt_N_results} 
to the form \Eq{bt_N_fit}.\label{bt_N_fit_results}}
\end{table}

\section{Summary and discussion}
\label{summary}

In this paper we have studied the validity of large-$N$ reduction for the
four dimensional quenched Eguchi-Kawai model. This model is a variant
of the original Eguchi-Kawai model in which the distribution of
the eigenvalues of the link matrices is forced to be uniform by
quenching, while all other degrees of freedom remain dynamical.

We find that while enforcing a uniform eigenvalue distribution
is indeed a necessary condition for large-$N$ reduction to hold, it is
not sufficient. The reason is that quenching fixes the
eigenvalues only up to permutations that can be performed independently in the
four directions. These permutations occur dynamically in the
model due to fluctuations in the unquenched degrees of freedom,
and can lead to correlations between the ordering of the 
eigenvalues of the four link matrices. If such correlations occur then 
we show that the
arguments of Refs.~\cite{BHN1,Migdal,HN,GK,Parisi-papers,DW}
for the validity of the large-$N$ quenched reduction break down.

The question then is whether such correlations between link eigenvalues
occur. We show that they are indeed expected in the weak-coupling regime by minimizing the free energy 
with respect to the ordering of the eigenvalues. This then
leads us to perform a detailed numerical study of the QEK model with
intermediate and strong couplings using Monte-Carlo techniques. We
find the weak-coupling calculation is indeed a good guide and obtain
the following evidence for the breakdown of large-$N$ reduction in the
model:
\begin{itemize}
\item We observe clear evidence for eigenvalue correlations by measuring order
parameters that explicitly probe the correlation between the different link matrices along the different Euclidean directions.
\item When we compare the plaquette expectation values of
the QEK model and of large volume lattice gauge theories, we find very
large discrepancies that do not go away with increasing $N$.  
\item
When we measure the coupling at which a strong-to-weak transition
occurs in the QEK model, and compare it to the coupling at which the
``bulk'' transition takes place in large-$N$ lattice gauge theories in
large volumes, we observe a large discrepancy which is of order
13\%, and very significant statistically.  
\end{itemize}
We checked that these conclusions
are insensitive to the precise form of
the quenched eigenvalue distribution,
and to the way we perform the quenched average.
We also considered values of $N$ up to $200$ to look for a late
onset of $1/N$ behavior, but find none.
We conclude that
the momentum quenched large-$N$ reduction of $SU(N)$ lattice
gauge theories fails in the continuum limit.

We have focused in this paper on the behavior in the weak coupling region,
  since this is where a continuum limit might be taken.
  Nevertheless, it is also interesting to consider the status of
  reduction in the strong coupling regime.
  In the strong-coupling expansion no eigenvalue correlations appear and so the QEK model is expected to be equivalent to the $SU(\infty)$ gauge theory for large enough `t Hooft coupling $\lambda$. It follows
  that reduction is valid until a transition occurs into a phase in
  which eigenvalue correlations appear. For the quenched Eguchi-Kawai this occurs at the strong-to-weak transition.
  We have checked numerically that the eigenvalue correlations do vanish on the strong-coupling side of this
  transition.
  A similar picture holds for both the EK and TEK models: reduction holds
  for large enough $\lambda$ but is
  lost below a certain coupling. We stress,
  however, that this transition coupling differs for all three theories (and
  also differs from the bulk transition coupling for $SU(\infty)$). This is
  just a reflection of the fact that the weak-coupling phases in these
  theories are unrelated. 

These results, together with those of Refs.~\cite{TV,Ishikawa} 
concerning the TEK model,
mean that, currently, only two single-site models are known
that can possibly
reproduce the properties of QCD at large-$N$.  The first is 
the ``deformed'' Eguchi-Kawai (DEK) model, which is the single-site
example of a class of models
proposed very recently in Ref.~\cite{DEK}.
In the DEK, the action of the Eguchi-Kawai model is deformed so that breakdown of the
$Z_N^4$ symmetry is energetically disfavored, and yet at the same time the large-$N$ dynamics is not modified.
Thus in this model the original Eguchi-Kawai proof of reduction
is expected to remain valid. In preliminary calculations of the DEK model we see that this deformation must include terms that decorrelate the gauge fields in different Euclidean directions, and this makes a direct connection to our results in the QEK model, where we see that such correlation is dynamically preferred.
Deformation comes, however, at a cost. Adding all possible deformations
is likely to be prohibitively expensive, because there are $\sim N^4$
in four dimensions. Whether one can improve this scaling by wise choices
of the deformations is a subject for future investigation.

 The other single-site candidate is the model obtained by adding $1\le
 N_f\le 4$ Majorana adjoint quarks, with periodic boundary conditions,
 to the Eguchi-Kawai action \cite{QCDadjoint}. Here the one-loop
 potential for the link eigenvalues is repulsive if the quark
 mass in units of the lattice spacing, $a_{\rm lat} m$, is small enough, and
 reduction is expected to hold. By taking $a_{\rm lat}m\ll 1$ and yet $m$ much
 larger than the confinement scale $\Lambda$, this single site model
 should reproduce the pure Yang-Mills theory in the IR, and it remains
 to be seen how such a construction compares computationally to that
 of \cite{DEK}. For $m\ll \Lambda$ this model describes the large-$N$
 limit of QCD with adjoint quarks which, for $N_f\le 4$, is expected to
 confine and to be related to $3$-color QCD through the orientifold
 large-$N$ equivalence. For $N_f\simeq 5$ this model is expected to be
 close to conformal, and also of interest. We leave the exploration of
 both these single-site models to future studies.

\acknowledgements{We thank Herbert Neuberger,
Mike Teper,  Mithat Unsal, and Larry
Yaffe for useful discussions, and especially Helvio Vairinhos who also
provided us with his TEK code, on which our
code was initially based. This study was supported in part by
the U.S. Department of Energy under Grant No. DE-FG02-96ER40956.}

\end{document}